\documentclass[fleqn,usenatbib]{mnras}

\usepackage{newtxtext,newtxmath}

\usepackage[T1]{fontenc}

\DeclareRobustCommand{\VAN}[3]{#2}
\let\VANthebibliography\thebibliography
\def\thebibliography{\DeclareRobustCommand{\VAN}[3]{##3}\VANthebibliography}

\usepackage{graphicx}	% Including figure files
\usepackage{amsmath}	% Advanced maths commands
\usepackage{amssymb}	% Extra maths symbols
\usepackage{ulem}       % delete text with lines

\newcommand{\Msun}{\,\ensuremath{\mathrm{M}_\odot}}
\newcommand{\revise}[1]{\textcolor{black}{#1}}

\newcommand{\hyphen}{\mathchar`-}

\title[{\it \lowercase{s}}-process in UFDs]{{\it \lowercase{s}-}process enrichment of ultrafaint dwarf galaxies}

\author[Y. Tarumi et al.]
{Yuta Tarumi$^{1}$\thanks{E-mail: yuta.tarumi@phys.s.u-tokyo.ac.jp},
Takuma Suda$^{2,3,4}$,
Freeke van de Voort$^{5,6}$,
Shigeki Inoue$^{7,8}$,
Naoki Yoshida$^{1,3,9}$,
\newauthor
Anna Frebel$^{10}$
\\
$^{1}$Department of Physics, School of Science, The University of Tokyo, Bunkyo, Tokyo 113-0033, Japan\\
$^{2}$Department of Liberal Arts, Tokyo University of Technology, Ota-ku, Tokyo 144-8535, Japan\\
$^{3}$Research Center for the Early Universe, School of Science, The University of Tokyo, Bunkyo-ku, Tokyo 113-0033, Japan\\
$^{4}$Open University of Japan, Mihama-ku, Chiba 261-8586, Japan\\
$^{5}$Max Planck Institute for Astrophysics, Karl-Schwarzschild-Straße 1, 85748, Garching, Germany\\
$^{6}$School of Physics and Astronomy, Cardiff University, Queens Buildings, The Parade, Cardiff CF24 3AA, UK\\
$^{7}$Center for Computational Sciences, University of Tsukuba, Ten-nodai, 1-1-1 Tsukuba, Ibaraki 305-8577, Japan\\
$^{8}$Chile Observatory, National Astronomical Observatory of Japan, Mitaka, Tokyo 181-8588, Japan\\
$^{9}$Kavli Institute for the Physics and Mathematics of the Universe (WPI), UTIAS, The University of Tokyo, Chiba 277-8583, Japan\\
$^{10}$Department of Physics and Kavli Institute for Astrophysics and Space Research, Massachusetts Institute of Technology, Cambridge, MA 02139, USA}

\date{Accepted XXX. Received YYY; in original form ZZZ}

\pubyear{2020}

\begin{document}
\label{firstpage}
\pagerange{\pageref{firstpage}--\pageref{lastpage}}
\maketitle

\begin{abstract}
% It should be a single paragraph not more than 250 words (200 words for Letters).
%No references should appear in the abstract.
%Recent spectroscopic observations of
%ultrafaint dwarf galaxies (UFDs) revealed that the small, old galaxies
%contain a substantial amount of neutron-capture elements such as
%Barium (Ba), Strontium (Sr), and Europium (Eu).
We \revise{study} the production of barium (Ba) \revise{and strontium (Sr)} in ultrafaint dwarf galaxies (UFDs). \revise{Both {\it r}- and {\it s}- processes produce these elements,}
and one can infer the contribution of the {\it r-}process  
from the characteristic {\it r-}process abundance pattern,
whereas the {\it s-}process contribution remains largely unknown.
We show that the current {\it s-}process yield from asymptotic giant branch (AGB) stars is not sufficient to explain the Ba \revise{and Sr} abundances observed in UFDs. \revise{Production of these elements} would need to be efficient from the beginning of star formation in the galaxies. The discrepancy of 
nearly \revise{or more than} $1$\,dex is not reconciled even if we consider {\it s-}process in super-AGB stars. 
We consider a possible resolution by assuming \revise{rotating massive stars (RMSs) and electron-capture supernovae (ECSNe) as additional contributors.} \revise{We find that the RMSs could be the origin of Ba in UFDs if $\sim 10$ per cent of massive stars are rotating at 300 km s$^{-1}$. As for ECSNe, we argue that their fraction is less than 2 per cent of core-collapse supernova. It narrows the progenitor mass-range to $\lesssim 0.1\,\Msun$ at $-3 \lesssim \mathrm{[Fe/H]} \lesssim -2$.}
We also explore another resolution by modifying the stellar initial mass function (IMF) in UFDs and find a \revise{top-light} IMF model that reproduces the observed level of Ba-enrichment.
Future observations that determine or tightly constrain the europium \revise{and nitrogen} abundances
are crucial to identify the origin of Ba \revise{and Sr} in UFDs. 
\end{abstract}


%%%%%%%%%%%%%%%%%%%%%%%%%%%%%%%%%%%%%%%%%%%%%%%%%%

%%%%%%%%%%%%%%%%% BODY OF PAPER %%%%%%%%%%%%%%%%%%

\section{Introduction}

Neutron-capture processes are important for the synthesis of the heaviest elements. There are two types, the \revise{{\it r-} and {\it s-}process}, depending on whether neutron-capture reactions occur faster than $\beta$-decays \citep[see, e.g.][]{2021Cowan_RprocessReview}.
Neutron-rich environments are necessary for neutron-capture reactions to take place, but the origin or the astrophysical production site remains debated for the {\it r-}process \citep{2021Cowan_RprocessReview, Nishimura17_iprocess}.
On the contrary, it is generally \revise{accepted} that the {\it s-}process occurs mainly in low and intermediate mass stars (2-8\,M$_{\odot}$) during their asymptotic giant branch (AGB) phase.

\revise{A major concern related to {\it s-}process nucleosynthesis is the exact site of neutron production. It is thought to occur above the helium core within thermally pulsing AGB stars during their interpulse phases.
It is assumed that protons in the convective envelope are mixed down into the upper layer of the helium core to trigger $^{13}\mathrm{C}(\alpha, n)^{16}\mathrm{O}$ reactions. These layers are called $^{13}$C pockets \citep[see, e.g.][]{Busso1999}.
The efficiency of the s-process nucleosynthesis by $^{13}$C pockets relies on free parameters, but it is chosen to reproduce consistent abundance trends for neutron-capture elements
\citep{2004Travaglio_sprocess, 2011Kappeler_sprocess_review}. Another site of neutron production is in convective shells driven by helium shell flashes in intermediate-mass AGB stars where neutrons are produced via $^{22}\mathrm{Ne}(\alpha, n)^{25}\mathrm{Mg}$ reactions \citep{Iben1975}.}

\revise{Other potential sites of {\it s-}process nucleosynthesis have also been explored by studying the signature of neutron-capture elements observed in metal-poor stars in the Galactic halo.
\citet{Suda2004} argue that neutrons are produced in the helium-flash convective zones when the convective shell reaches the bottom of the hydrogen-burning shell in low-mass AGB stars at extremely low-metallicity.
This mechanism could be the source of {\it s-}process elements in carbon-enhanced metal-poor stars. Fast-rotating massive stars are also considered in order to explain the abundance of {\it s-}process elements in these stars \citep{Choplin2017}.
}

\revise{Theoretical models that calculate {\it s-}process nucleosynthesis for different astrophysical sites can be tested with observed neutron-capture elements in metal-poor stars. In particular, barium (Ba) and strontium (Sr) are often used as characteristic tracers.}
However, since the main contributors of the {\it s-}process are 
long-lived, lower-mass stars, they contribute to the Galactic chemical evolution only slowly. This feature is consistent with the observations of Milky Way (MW) stars which show monotonic increase of [Ba/Fe] as the iron (Fe) abundance [Fe/H] increases. 

To learn in more detail about the various environments in which s-process(es) may occur and how so, it is useful to consider dwarf galaxies. They have a considerable advantage over halo star samples in that they 1) can be simulated in its entirety (e.g. \citealt{2021Tarumi_UFDmerger}), and 2) observations of stars in surviving systems can be used to contrast those simulations (e.g. \citealt{2019Brauer_disrupted_dwarf}). Finally, comparisons can also be made with halo star results. 

Especially {ultra-faint dwarfs (UFDs)} have several distinct features in terms of their chemical evolution, which make them particularly interesting objects to study \citep{Simon19_UFDreview}. One important feature is the predominantly old stellar population. It is theoretically expected that the shallow gravitational potential well is inefficient at retaining gas after cosmic reionization \citep{2014Brown_UFDquench}. \revise{The quenching due to reionization is also confirmed in cosmological UFD simulations \citep{2017Jeon, 2019Wheeler, 2020Agertz}.} Hence, UFDs are chemically less evolved systems. Detailed results thus offer important insight into the early chemical evolution in the Universe. Another important trait is the stochasticity with which individual enrichment events occur. Given their low gas mass, rare enrichment events cause particularly strong signatures in any subsequently formed stars which can be modelled. In turn, the absence of any such signatures readily implies a deficit of element production.

By now, detailed chemical abundances have been obtained for metal-poor stars in many UFDs (e.g. \citealt{2014Frebel_BaFe, 2014Ishigaki_BaFe, 2018Chiti_BaFe, 2019Ji_GCandUFD}). These observations consistently show that neutron-capture element abundance (Sr and Ba) are low, and among the very lowest found in equivalent metal-poor halo stars. This suggests that the environments in early UFDs was likely different than those from which typical halo stars formed, which is likely larger dwarf galaxies. But exceptions exist also; three galaxies (out of $\sim 15$ UFDs) unambiguously show signatures of the r-process \citep{Ji16_RetII, 2020Hansen_GrusII}, as evidenced by pronounced europium (Eu) lines that could be detected (along with Sr and Ba) in stellar spectra. 

In the absence of Eu line detections or only low or moderate Sr and Ba enhancements, it becomes challenging to differentiate which process may have produced the observed abundances. After all, both the s- and r-process can produce Sr and Ba \citep{2006Cescutti_MW_sprocess, 2019Hirai_Sr_in_dwarfs}. As a consequence, it is interesting to explore the chemical enrichment history of those UFDs that did not experience a rare and prolific r-process event to understand to what extent s-process contributions can explain the low observed neutron-capture element abundances. 

In the present paper, we thus model the {\it s-}process enrichment of UFDs and contrast our results with observations for r-process UFDs. We use highly resolved cosmological simulations to follow the star formation and chemical enrichment of several systems, and our technical details are outlined in Section 2. Given the characteristic star formation histories and the small stellar masses of UFDs, we can derive clues towards understanding the chemical enrichment history of neutron-capture elements.
We present these simulation results in Section 3. In Section 4, we argue that the existence of an additional Ba source, such as rotating massive star (RMS), is strongly favoured to explain the observed elemental abundances in UFD stars.

\section{Method}

\subsection{Cosmological Simulations}
\begin{figure}
    \centering
    \includegraphics[width=\columnwidth]{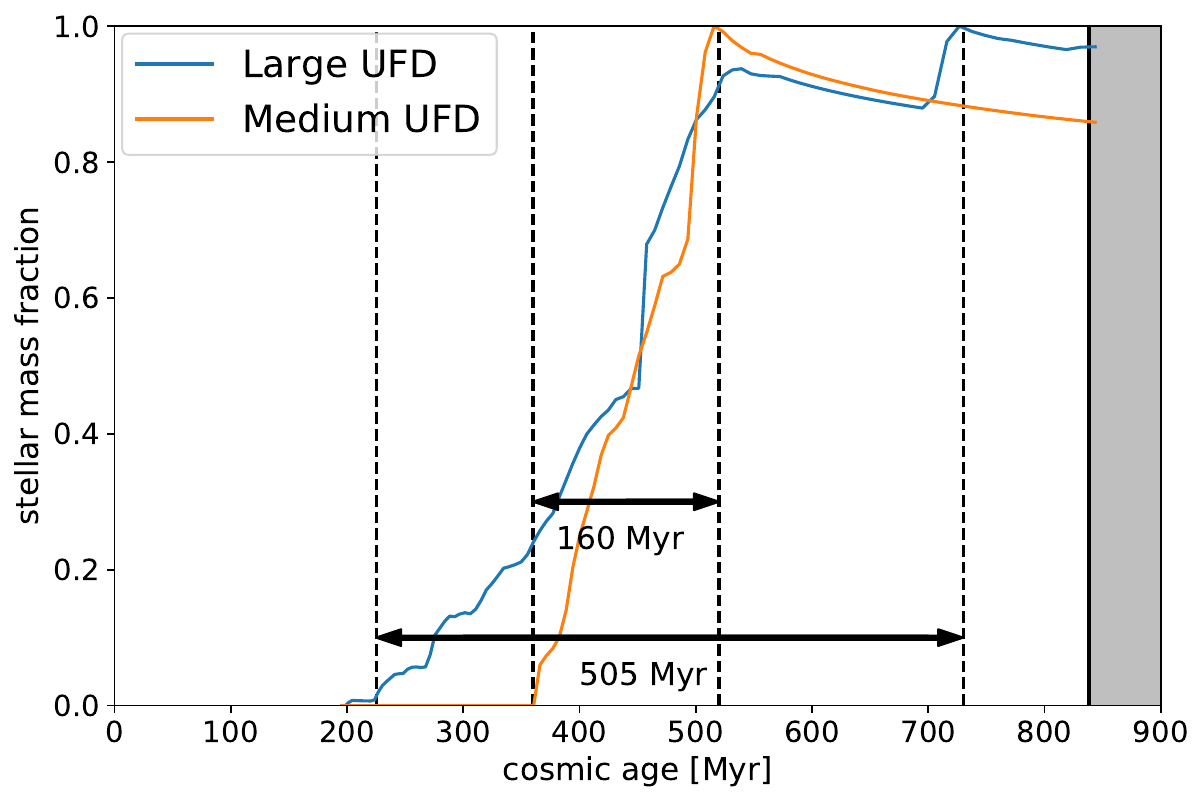}
    \caption{Star formation histories of our simulated galaxies. Blue curve: L-UFD, which forms stars for $\sim 500$\,Myr. The peak stellar mass is $\sim 12,000$\,\Msun. Orange curve: M-UFD, which forms stars for $\sim 160$\,Myr. The peak stellar mass is $\sim 4000$\,\Msun. The L-UFD experienced a major merger at a cosmic age of $\sim 500$\,Myr at redshift 10. The shaded area shows when the Universe became ionized and star formation was largely quenched.}
    \label{fig:star formation histories}
\end{figure}

We use \textsc{arepo}, a moving-mesh hydrodynamic simulation code \citep{Springel10_AREPO, Pakmor16_AREPO_improvement, 2019AREPOrelease}. As the cosmological parameters, we use the Planck 2018 results \citep{Planck2018}: $\Omega_{m} = 0.315, \Omega_{b} = 0.049, \sigma_{8} = 0.810, n_{s} = 0.965, H_{0} = 67.4 \mathrm{km s^{-1}Mpc^{-1}}$. The initial conditions are generated using \textsc{music} \citep{Hahn11_MUSIC}. The box-size is 1 comoving $h^{-1}$Mpc on a side. We use a hierarchical zoom-in technique to resolve the inner structures of the simulated galaxies. The mass of each dark matter particle is about $100\,\Msun$, and the typical mass of each gas cell and star particle is about 20\,\Msun. 
\revise{Our sample galaxies are the same as the ones we used in our previous paper \citep{2020Tarumi}.}
In this paper, we refer to halos 1, 2, and 3 as the large, medium and small UFDs (hereafter `L-UFD', `M-UFD', and `S-UFD'). \revise{The stellar mass of the S-UFD is $\sim 100\,\Msun$, and the number of the star particles is too small to make a robust comparison with the observations. Therefore, we mainly discuss the results from the L-UFD and M-UFD.} In Fig.~\ref{fig:star formation histories}, we present the star formation histories of the three galaxies. Halo masses ($M_{h}$), stellar masses ($M_{*}$), and sizes ($R_{200}$) of these samples are ($M_{h}, M_{*}, R_{200}$) = ($1.7\times 10^{8}\,\Msun$, $12,000\,\Msun$, $1.7$ kpc), and ($1.0\times 10^{8}\,\Msun$, $4,100\,\Msun$, $1.6$ kpc) at redshift $z=8$, which can be regarded as UFD progenitors \citep{Safarzadeh18_UFDselection}. These haloes grow to ($M_{h}, M_{*}, R_{200}$) = ($2.1\times 10^{8}\,\Msun$, $12,900\,\Msun$, $2.5$ kpc), and ($2.0\times 10^{8}\,\Msun$, $3,800\,\Msun$, $2.2$ kpc) by redshift $z=6.4$ which is our final snapshot. 

\revise{We do not adopt the ISM model typically used in \textsc{arepo} simulations \citep{Springel03_WindParticles}. Instead, we follow gas cooling and heating self-consistently 
as in \citep{2019Inoue_ISM} (see their 'single-phase ISM model').
The smooth gas density field is represented with Voronoi cells. Each gas cell contains physical properties such as density or temperature. Spatial resolution is progressively increased to suppress artificial fragmentation 
owing to insufficient resolution \citep{1997Truelove_TrueloveCondition}. \citet{2010Ceverino_Ncells} suggest that the Jeans length $\lambda_\mathrm{J} = \pi c_\mathrm{s}/G\rho_\mathrm{g}$ should always be resolved at least with seven cells to obtain converged result. In our simulations, we impose a non-thermal pressure floor to satisfy the condition proposed by \citet{2010Ceverino_Ncells}. We set the smallest scale for the pressure support at 0.2 comoving pc $h^{-1}$.}

\revise{
The star-formation rate (SFR) of each gas cell is calculated as $\mathrm(SFR) = 0.079m_\mathrm{g}/t_\mathrm{SF}$, where $m_\mathrm{g}$ is the mass of the gas cell and $t_\mathrm{SF} = (G\rho_\mathrm{g})^{(-1/2)}$ is the star formation timescale. We allow star formation only in gas cells with $\rho_\mathrm{g} > \rho_\mathrm{th} = 100\ \mathrm{atom\ cm^{-3}}$ and temperatures $T < 10000\ \mathrm{K}$. }

\revise{Stellar feedback is modelled by ejecting `wind particles', which model the outflow driven by type-II SNe. Gas cells satisfying the SF conditions stochastically form stars or wind particles. The ratio between the numbers of star particles and wind particles is determined from energy deposition by type-II SNe, assuming $8-100\,\Msun$ of stars explode immediately. Each SN deposits $1.7\times 10^{51} \mathrm{erg}$ in the form of wind particles. The number of SNe that occur in each star particle is determined stochastically so that the number of events is always integer. These wind particles carry 40 per cent of the original metal contents \citep{Vogelsberger13_MetalStripping}. Each newly formed wind particle is ejected into a random direction with the initial velocity of $3.46 \sigma_\mathrm{DM}$, where $\sigma_\mathrm{DM}$ represents the one-dimensional velocity dispersion of the ambient dark matter particles \citep{Springel03_WindParticles}. After the ejection, the orbits of wind particles are calculated purely gravitationally. They keep travelling until they reach a low-density gas cell with $\rho_\mathrm{g} < 0.05 \rho_\mathrm{th}$ or the maximum lifetime. A disappearing wind particle deposits its mass, metals, momentum, and energy into the gas cell it reached. Energy deposition of type-Ia SNe and AGB stars are not taken into account.}

\subsection{Chemical evolution of ultrafaint dwarf galaxies}

\revise{Each star particle in our simulations represents a stellar cluster with a population expressed by the Chabrier's IMF \citep{Chabrier01_IMF}.}
It enriches the surrounding gas cells with metals.
\revise{For type-II SNe,} the amount of elements distributed to nearby cells in one time-step are determined as follows. First, we specify the mass range of stars that contribute to the chemical enrichment within the next time-step using the age and metallicity of the star particle. Next, we integrate the elemental yields from the stars that die within the time-step. \revise{We use metallicity-dependent type-II SNe yields and lifetimes obtained from \citet{1998Portinari_typeIIyield}.}
Finally, we distribute the calculated amounts of elements to nearby gas cells 
in the smoothed-particle-hydrodynamics manner. \revise{The kernel size is determined so that effectively  $64 \pm 1$ gas cells are involved in the metal enrichment from each star particle.} The procedure is performed for each star particle at each time-step.
\revise{Type-Ia SNe only occur with a delay after their star formation. We use the empirical power-law delay-time distribution (DTD) by \citet{2012Maoz_Ia}. The minimum delay-time is set at 40\,Myr, the power is $-1.12$, and 0.0013 type-Ia SNe occur per $1$\,\Msun\ of stars formed within 10\,Gyr. We utilize yields from \citet{2004Travaglio_typeIayield}.}
\revise{The median [Fe/H] of L- and M-UFDs are $\mbox{[Fe/H]}=-2.66$ and $\mbox{[Fe/H]}=-2.96$. Since [Fe/H] of UFDs are $-3$ to $-2$ (e.g. $-2.71$ for Segue 1; see \citealt{Simon19_UFDreview} for the list of UFDs).
Our galaxy samples reproduce the [Fe/H] ranges of observed UFDs.}

The masses of star particles in our high-resolution simulations are too small for \revise{applying the single stellar population (SSP) approximation in which the elemental yield is calculated as the average of contributions from all the stellar mass range. This would lead to an artificial fractionalization of the number of stars. 
Therefore, we stochastically sample the number of AGB stars that contribute to {\it s}-process enrichment. We use the {\it s}-process yields from the FRUITY database \citep{2015Cristallo_FRUITY, 2016Cristallo_FRUITY}. {\it S}-process yields are metallicity dependent so we use yields consistent with the metallicity of the star particles. The lowest metallicity available is $\log(Z/Z_{\odot}) = -2.3$ which is similar to the metallicity of our UFDs. For {\it s}-process elements, yields are available for the following masses of 1.3, 1.5, 2.0, 2.5, 3.0, 4.0, 5.0, and 6.0\,\Msun. 
To calculate the number of stars with the mass of each grid, we take bins separated at the midpoint of each grid point, namely, 1.4, 1.75, 2.25, 2.75, 3.5, 4.5, 5.5\,\Msun. The first bin starts from 1.0\,\Msun, and the highest bin ends at 6.5\,\Msun. The exact choice does not affect the results because stars with the min/max masses do not contribute significantly.
We then calculate the expected number of stars within each of these bin assuming a Chabrier IMF \citep{Chabrier01_IMF} for each star particle.}

In our \revise{fiducial} model, Sr is produced only in AGB stars
although various channels for Sr production have been proposed. One is electron-capture supernovae (ECSNe) (see e.g. \citealt{2011Wanajo_ECSNe}). \revise{Unfortunately, the mass range of ECSNe is not well constrained, and the choice of range greatly affects the results. 
We thus treat the Sr production channel as an additional source, and discuss its effects separately. We describe how we treat the additional sources in the next subsection.}

\subsection{Additional sources of {\it s}-process elements}

\revise{We take two non-standard origins into account: rotating massive stars (RMSs) and ECSNe. We include these sources separately and discuss each effect individually. We call these models as `RMS model' and `ECSN model'. In our RMS model, {\it s}-process element synthesis can occur if rotation mixes the various layers in a star.
Slowly rotating stars preferentially produce lighter elements, such as Sr, while fast rotating stars produce copious amounts of heavy elements such as Ba. We adopt yields from \citet{2018Limongi_Chieffi_RMS}. The initial distribution of rotation velocities (IDROV)} is adopted from \citet{2018Prantzos_RMS}.

\revise{Stars with masses of the lowest end of CCSN mass-range explode as ECSNe \citep{1987Nomoto_ECSNe, 2008Janka_ECSNe, 2018Wanajo_ECSNe}. In ECSNe, weak {\it r}-process can occur in the centre region of the star and produce a copious amounts of Sr. SNe 2018zd and the origin of Crab pulsars might be the result of such an ECSNe \citep{2020Hiramatsu_might_be_ECSNe}, but until now there has been no observational confirmation on an ECSNe explosion. As we cannot determine the fraction of ECSNe out of all CCSNe, the mass range remains uncertain, and typically 0-1$\,\Msun$ is assumed. We assume $7.9\times 10^{-5}\,\Msun$ of Sr is produced, following the e8.8 model of \citet{2018Wanajo_ECSNe}. \citet{2019Hirai_Sr_in_dwarfs} constrained the mass-range of ECSNe to be $8.2$-$8.4\,\Msun$ by modelling stellar Sr abundances in dwarf galaxies. In this work we assume that ECSNe occur every 5000\,\Msun\ of stars formed. This fraction corresponds to assuming that 1/50 of SNe are ECSNe, equivalent to assuming that $8.0$-$8.1\,\Msun$ stars explode as ECSNe.}

\section{Results}

\begin{figure*}
    \centering
    \includegraphics[width=\columnwidth]{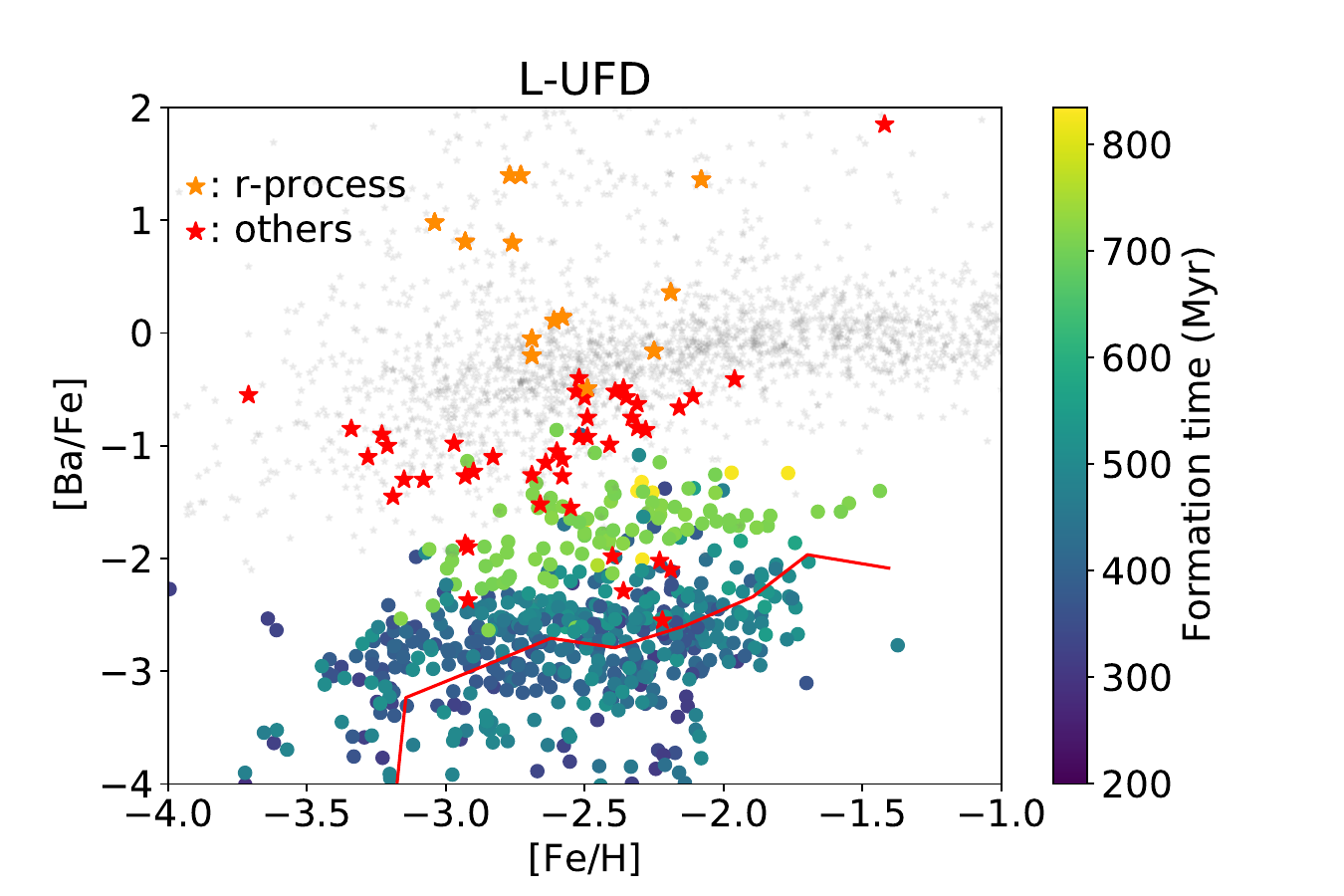}
    \includegraphics[width=\columnwidth]{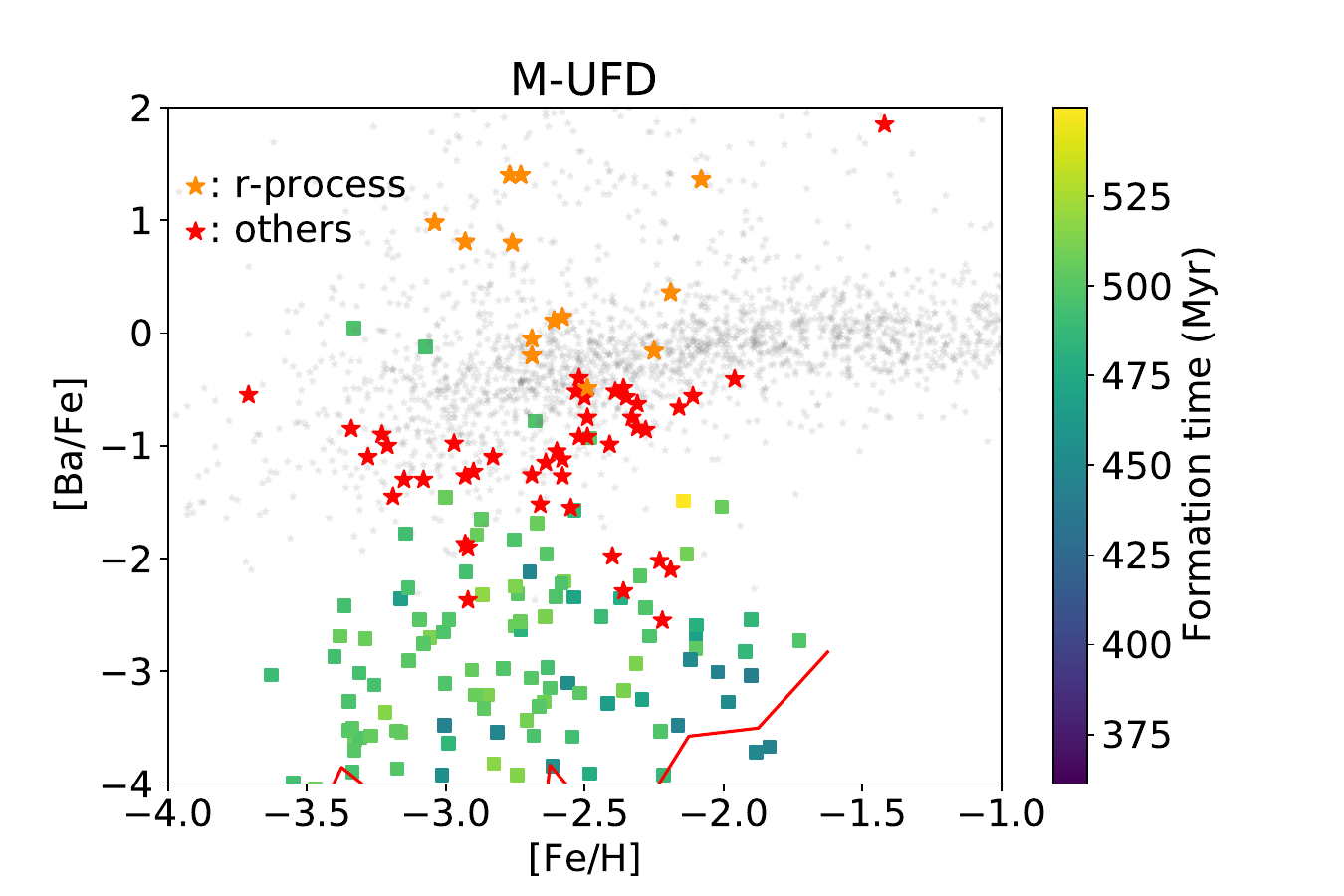}
    \includegraphics[width=\columnwidth]{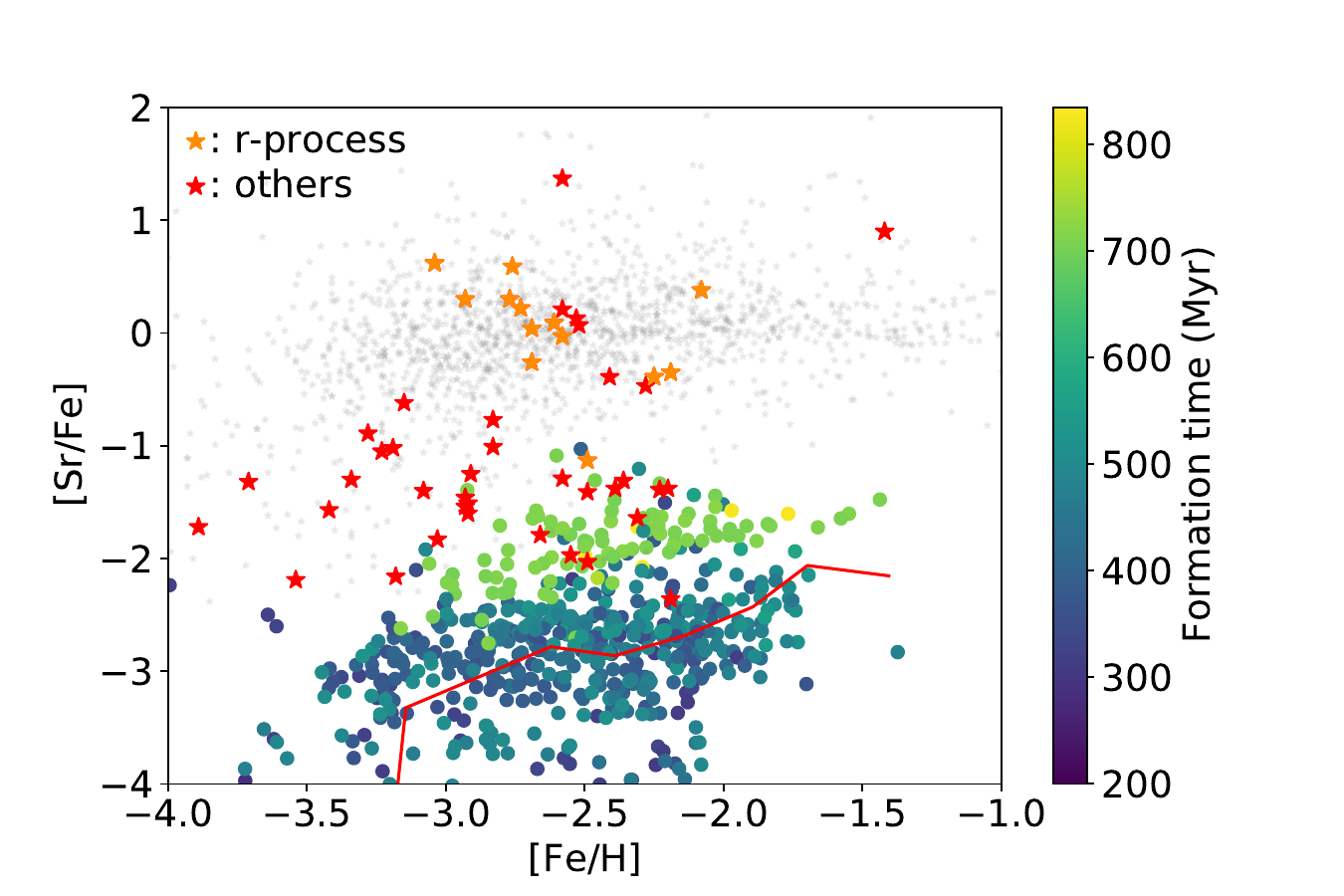}
    \includegraphics[width=\columnwidth]{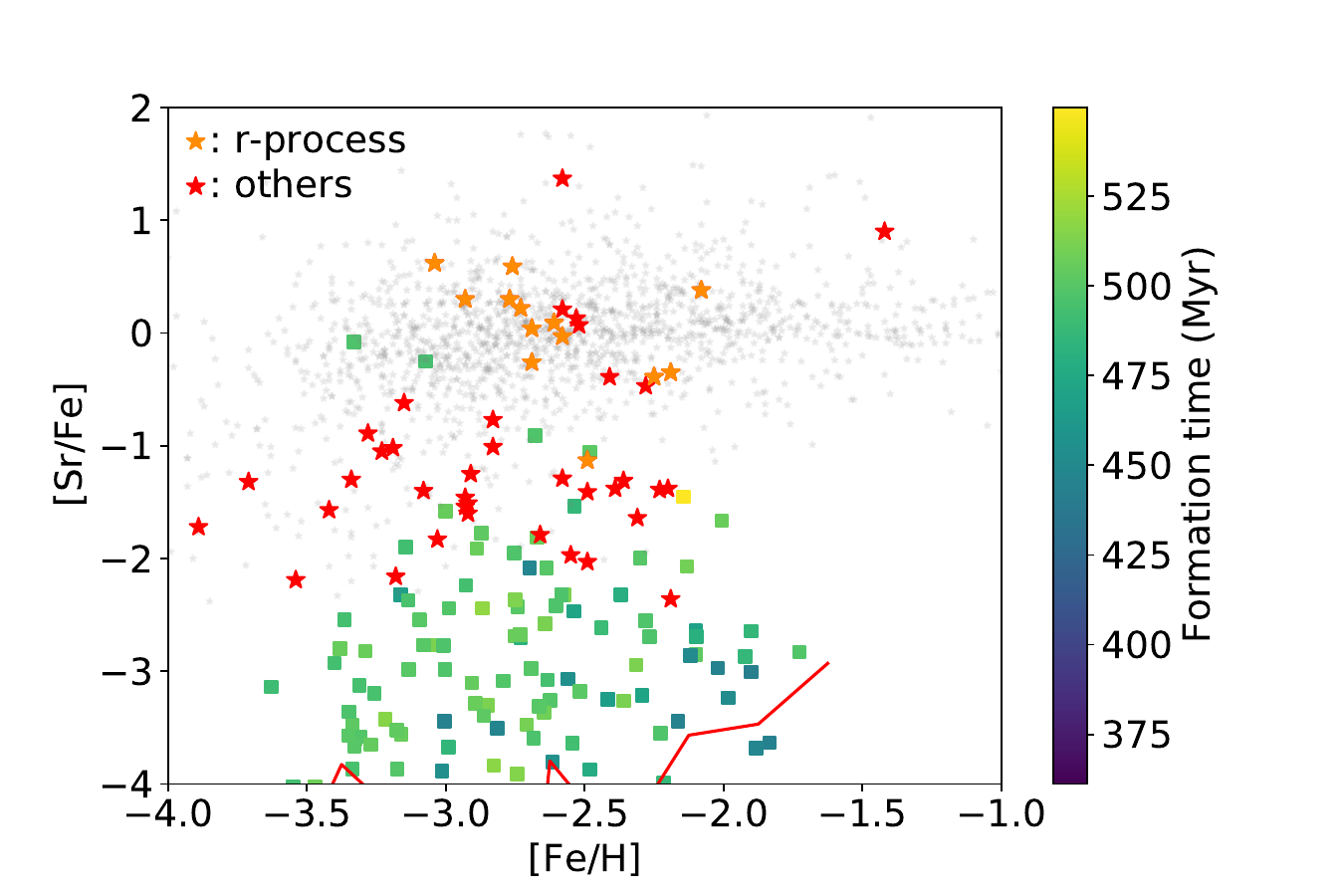}
    \caption{[Ba/Fe] (top panels) and [Sr/Fe] (bottom panels) as a function of [Fe/H] for our model stars in our two L- and M-UFD models (filled circles and squares, respectively; left and right panels, respectively) obtained at $z = 6.4$, and
    compared with observational data of stars in UFDs. Colour-coded is the age of stars. The stars that show the characteristic abundance patterns of the {\it r-}process (seven stars in Ret~II, five stars in Tuc~III, and one star in Grus~II) are plotted with orange star symbols. Other UFD stars are plotted with red star symbols. For reference, stars in the MW halo are plotted with grey dots. UFDs represented here are Bootes I, Bootes II, Canes Venatici II, Coma Berenices, Carina II, Carina III, Grus I, Hercules, Horologium I, LeoIV, Pisces II, Reticulum II, Segue 1, Segue 2, Triangulum II, Tucana II, Tucana III, Ursa Major, Ursa Major II, and Willman I. 
    The solid red lines depict the medians of [Ba/Fe] and [Sr/Fe] for each [Fe/H] 0.25\,dex bin for our model stars. 
    Observational abundances are obtained from the SAGA database \citep{2008SAGA, SAGA_dwarf}, together with the latest data added manually, based on the following studies \citep{2007Martin_BaFe, 2008Koch_BaFe, 2009Aden_BaFe, 2009Feltzing_BaFe, 2009Koch_BaFe, 2010Frebel_BaFe, 2010Norris_BaFe1, 2010Norris_BaFe2, 2010Norris_BaFe3, 2011Aden_BaFe, 2011Lai_BaFe, 2011Simon_BaFe, 2013Gilmore_BaFe, 2013Koch_BaFe, 2013Vargas_BaFe, 2014Frebel_BaFe, 2014Ishigaki_BaFe, 2014Koch_BaFe, 2016Francois_BaFe, 2016Ji_BaFe, 2016Ji_BaFe2, 2016Ji_RetII_ApJ, 2016Martin_BaFe, 2016Roederer_BaFe, 2017Hansen_TucIII, 2017Kirby_BaFe, 2017Venn_BaFe, 2018Chiti_BaFe, 2018Ji_BaFe, 2018Nagasawa_BaFe, 2019Ji_BaFe, 2020Hansen_GrusII, 2020Ji_CarII}.
   } 
    \label{fig:abundance scatter plot}
\end{figure*}

%===========================================
\subsection{AGB {\it s}-process enrichment model}

In the fiducial model, {\it s}-process element production occurs only in AGB stars. The yields table is obtained from the FRUITY database \citep{2015Cristallo_FRUITY, 2016Cristallo_FRUITY}. We do not consider {\it r}-process enrichment. Here we compare our simulated results to UFD stars with Ba or Sr detections. There are some observations only with upper limits. We do not consider these stars for now since most upper limits are meaninglessly high due to poor data quality, but discuss implications further below.

The abundances of {\it s}-process elements show large variation even among our simulated UFDs, which are uncontaminated by {\it r}-process enrichment. Ba and Sr abundances are higher in the L-UFD than in \revise{M- and S-} UFDs, even if we compare stars with similar Fe abundances. The different abundances reflect the difference in the star formation histories. \revise{The L-UFD forms stars for 500\,Myr, whereas the M- and S- UFDs cease their star formation after 150\,Myr from their first stellar production.}
In galaxies with long star formation duration, elements that low-mass stars synthesise are incorporated only into stars that form much later. Therefore, the abundances of {\it s-}process elements are higher in the L-UFD due to the larger contribution from low-mass stars.

Fig.~\ref{fig:abundance scatter plot} shows abundance ratios of  neutron-capture elements [Ba/Fe] and [Sr/Fe] as a function of [Fe/H] for stars observed in 20 UFDs. Overlaid are our simulation results from the L- and M-UFDs that are obtained at $z=6.4$. \revise{We omit results from S-UFD since very little {\it s}-process elements are incorporated into stars to meaningfully contribute anything.} \revise{The epoch of star formation scales with symbol colour.}

UFDs can be classified into two groups by whether the {\it r-}process has had a dominant, measurable effect on their chemical evolution. Reticulum~II (Ret~II), Tucana~III (Tuc~III) and Grus~II (Gru~II) are the ones with clear signatures of the {\it r-}process\footnote{There is one star with a significant Eu and Ba enhancement in Segue 1. It is a CH star, and the high abundance of the neutron-capture elements is thought to originate from mass transfer from the former AGB star in a binary system and not by previous neutron-capture processes \citep{2014Frebel_BaFe}.}. 
Stars in these UFDs (plotted with orange star symbols) show systematically higher Ba and Sr abundances than those in other UFDs (plotted with red star symbols), commensurate with expectations of a prolific r-process enrichment event. The large amounts of Ba and Sr produced by the {\it r-}process event would overpower signatures left over from any neutron-capture processes that may occur before or after. 

But to make matters more complication, at least two stars in Ret II display rather low Sr and Ba values (not shown in the figure but see Fig. 3 in \citealt{2016Ji_RetII_ApJ}), and the UFDs without clear {\it r}-process enrichment do contain a lot of stars with significant amount of {\it s}-process material. Nevertheless, our results clearly show that for stars with measurable neutron-capture element abundances AGB stars alone are not sufficient source to explain the origin of Ba and Sr for these stars. However, we also note that likely more stars exist in these systems that have lower Sr and Ba measurements than what is represented with the current observed sample\footnote{Although there are stars  with upper limits we chose to not shown them in the figure since the limits are largely very high and thus meaningless, owing to insufficient data quality in these often very faint stars. This renders it difficult, or often even impossible, to detect weak, blue lines such as Sr\,II at 4077\,{\AA}.}. Those stars 1) might not yet have such measurements due to being faint and corresponding poor data quality especially in the blue spectral regime preventing even meaningful upper limits, 2) or these stars simply have not even been observed yet at all. Such a putative low Sr-Ba-population may well be explained solely with an AGB origin scenario. Future observations can hopefully shed further light on this issue so that our predictions could be further put to the test. In fact, there are hints of this being the case. Metal-poor halo stars have [Ba/Fe] going down to about $\mbox{[Ba/Fe]}\sim-2.0$ which is roughly at the level of the simulated stars formed last in our simulation (light green circles in the figure). Furthermore, all stars with red symbols with $\mbox{[Ba/Fe]}>-2.0$ form a branch not unlike that of Milky Way stars albeit offset to lower [Ba/Fe] values. The lower of these stars overlap with the top end of the simulated later-time stars -- clearly more overlap would be achieved with improved spectral line detections should additional data become available (e.g. with the Giant Magellan Telescope). 

\revise{The scatter of [Ba/Fe] and [Sr/Fe] in each galaxy is also an interesting piece of information. Observationally, Bootes I (Boo~I), Carina II (Car~II), and Coma Berenices (ComBer) show significant scatter while most other UFDs do not. We also note that the number of observations is limited due to observational difficulties such as stars being too faint to obtain higher signal-to-noise observations while also requiring high spectral resolution. Therefore, even for UFDs with  little apparent scatter, significant scatter in [Ba/Fe] and [Sr/Fe] may still be present within. In our simulation, large scatter is observed both in the L- and M-UFDs. The Ba content of S-UFD is too little, and therefore we cannot discuss the scatter. }

\revise{There are two kinds of inhomogeneities to diversify the abundances of the neutron-capture elements within one system, spatial and temporal inhomogeneities. The colours of the symbols in Fig.~\ref{fig:abundance scatter plot} show that the L-UFD is more affected by  temporal ones, while the M-UFD is more affected by spatial inhomogeneities. This behaviour is consistent with the star formation histories of these galaxies. The M-UFD forms stars within 160\,Myr 
which is too short to mix the ISM efficiently \citep{2020Tarumi}. The L-UFD forms stars for $\sim$\,500\,Myr, and therefore, stars are typically formed from well-mixed gas. However, instead the temporal-induced scatter becomes important which results large scatter in the neutron-capture element abundances.}

\subsection{Adding yields of rotating massive stars}

\begin{figure*}
    \centering
    \includegraphics[width=\columnwidth]{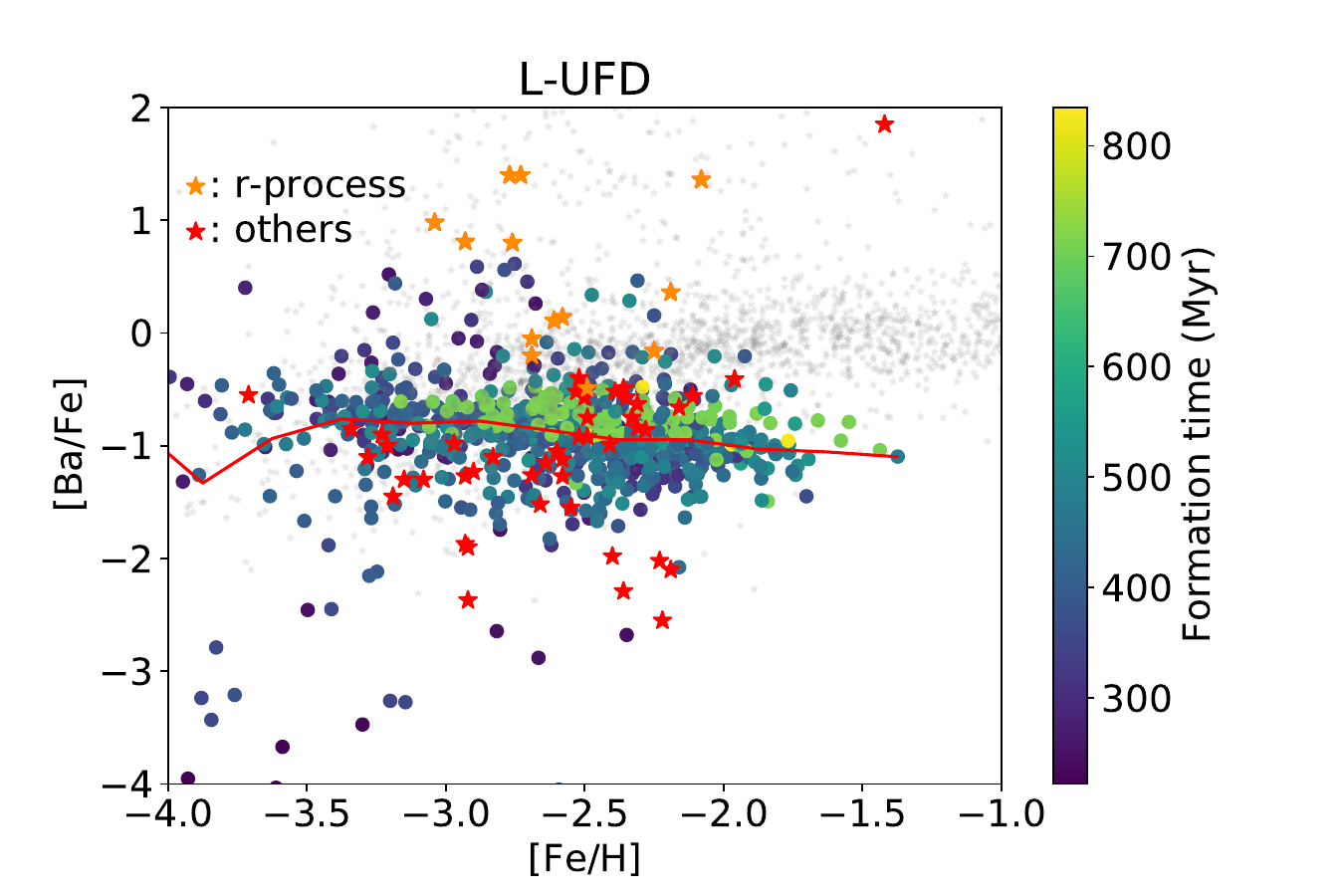}
    \includegraphics[width=\columnwidth]{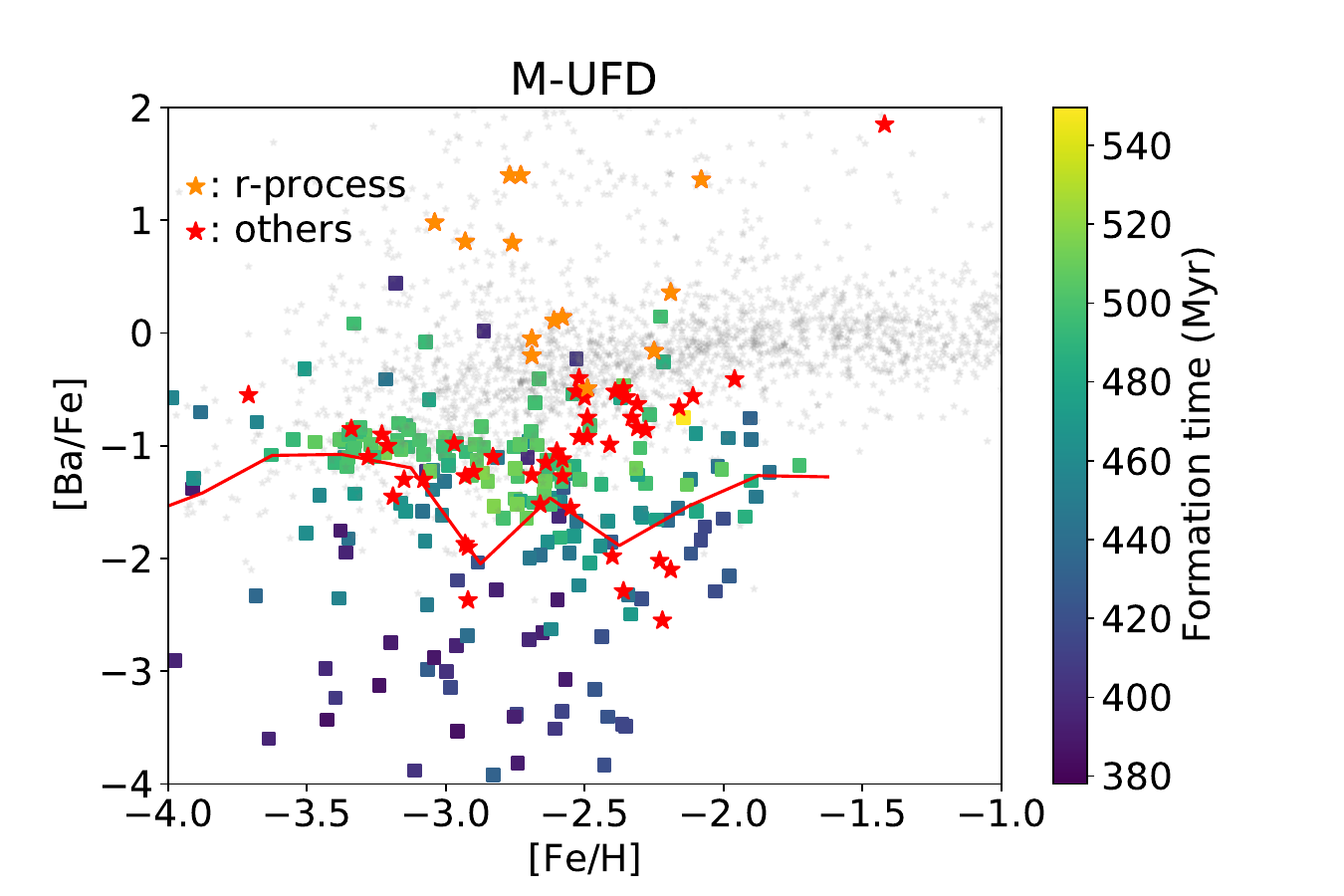}
    \includegraphics[width=\columnwidth]{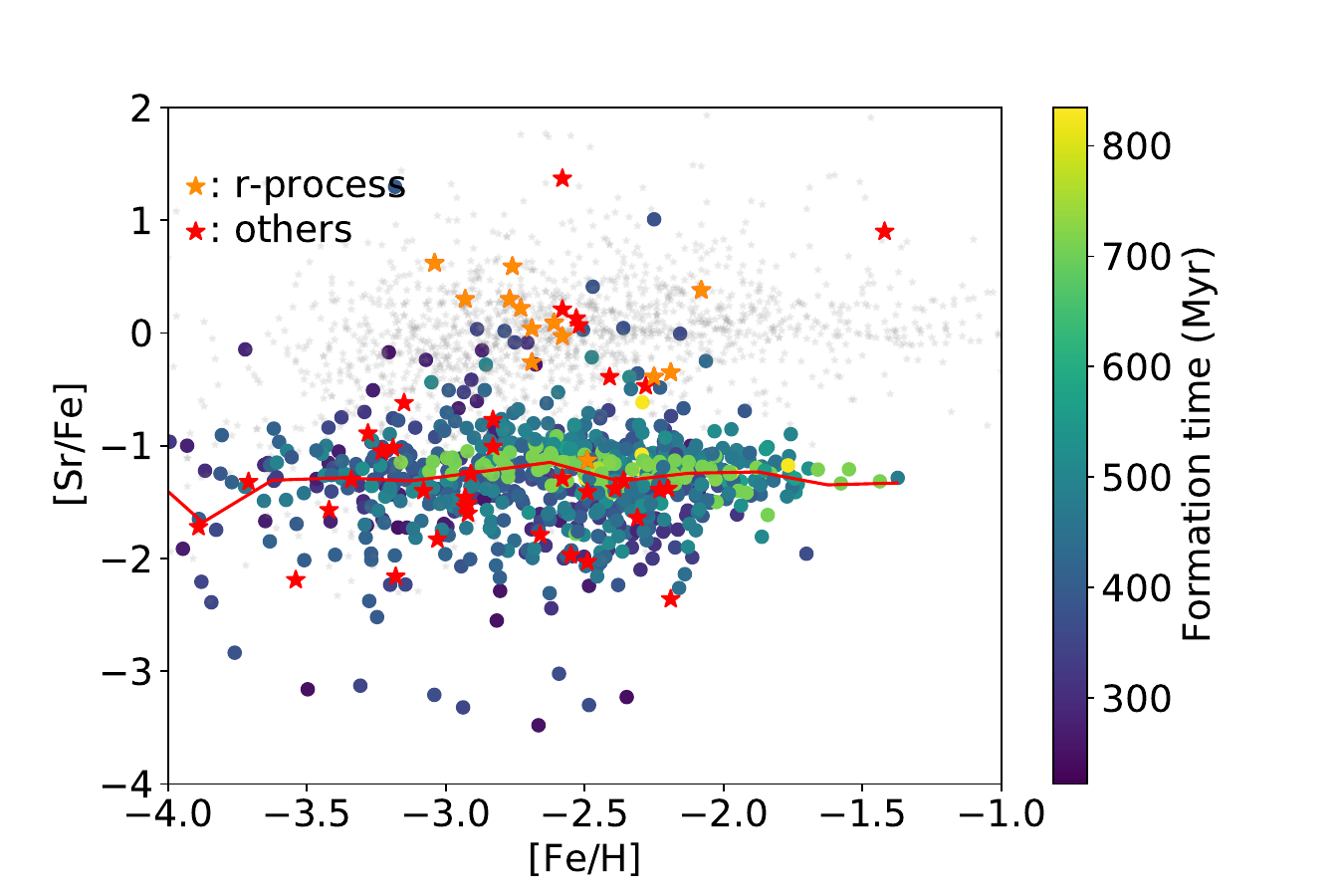}
    \includegraphics[width=\columnwidth]{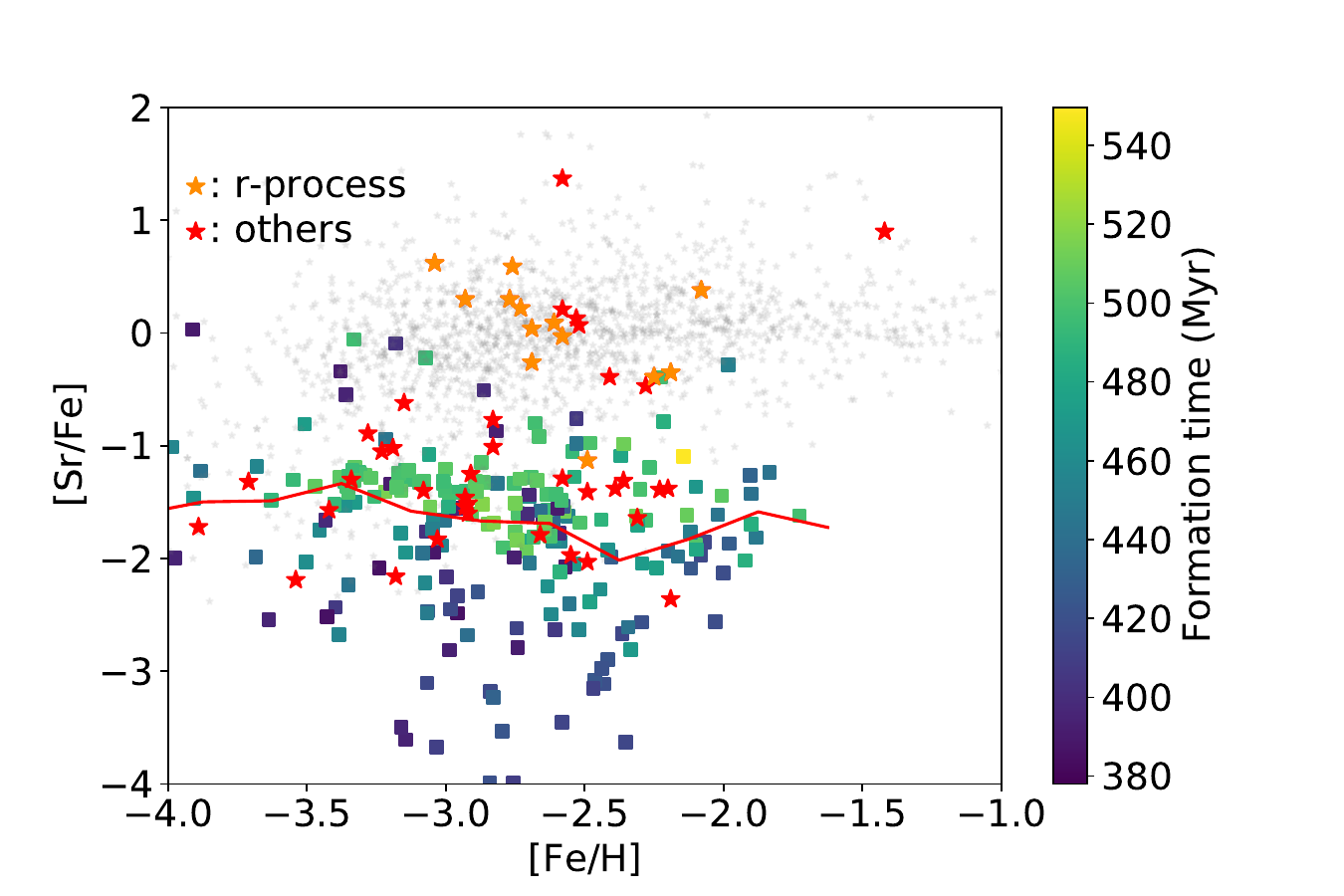}
    \caption{Observed abundances in UFDs overlaid with our RMS model results for [Ba/Fe] (top panels) and [Sr/Fe] (bottom panels). Symbols and colours are the same as in Fig. 2. The simulated abundances agree well with the observed abundances, including the largely flat trend with [Fe/H].
}    
    \label{fig:Rotating massive stars}
\end{figure*}

We test how our fiducial model results change when including an additional Ba and Sr source originating from RMS. We take yields from \citet{2018Limongi_Chieffi_RMS}. For the rotation velocity distribution, we assume that 23 and 2 per cent of stars are rotating at 300\,km\,s$^{-1}$ at metallicities $Z=-3\ \mathrm{and} -2\ Z_\mathrm{\odot}$, respectively, and 72 and 48 per cent of stars are rotating at 150\,km\,s$^{-1}$ at metallicities $Z=-3\ \mathrm{and}\ -2\ Z_\mathrm{\odot}$, respectively, following \citet{2018Prantzos_RMS}. 

The additional RMS contribution helps to reproduce the typical values of [Ba/Fe] and [Sr/Fe] observed in UFDs and an overall largely flat distribution with increasing [Fe/H]. In Fig.~\ref{fig:Rotating massive stars}, we show these results. 
This is strikingly different from our fiducial model where Ba and Sr originate from delayed and metallicity-dependent sources, and hence, [Ba/Fe] increases with [Fe/H]. 
However, in this RMS model, the dominant producer of Ba is massive stars rotating at 300 km s$^{-1}$. Therefore they have very short delay. The yield increases with metallicity, although the fraction of fast-rotating stars decreases with metallicity, and these effects compensate each other. We note that in some UFDs that show very low [Ba/Fe], Ba might originate from relatively rare stars, and that the low stellar mass (e.g. Segue 1) is not enough to sample such stars.

\revise{A relevant feature is the small abundance scatter in L-UFD. Since the RMS have no delay time, it erases temporal inhomogeneities, in both cases of [Ba/Fe] and [Sr/Fe]. This can be seen in the left panels in Fig.~\ref{fig:Rotating massive stars}. Note that the scatter is larger in the M-UFD and spatial inhomogeneities still exists in this system.}

\begin{figure*}
    \centering
    \includegraphics[width=\columnwidth]{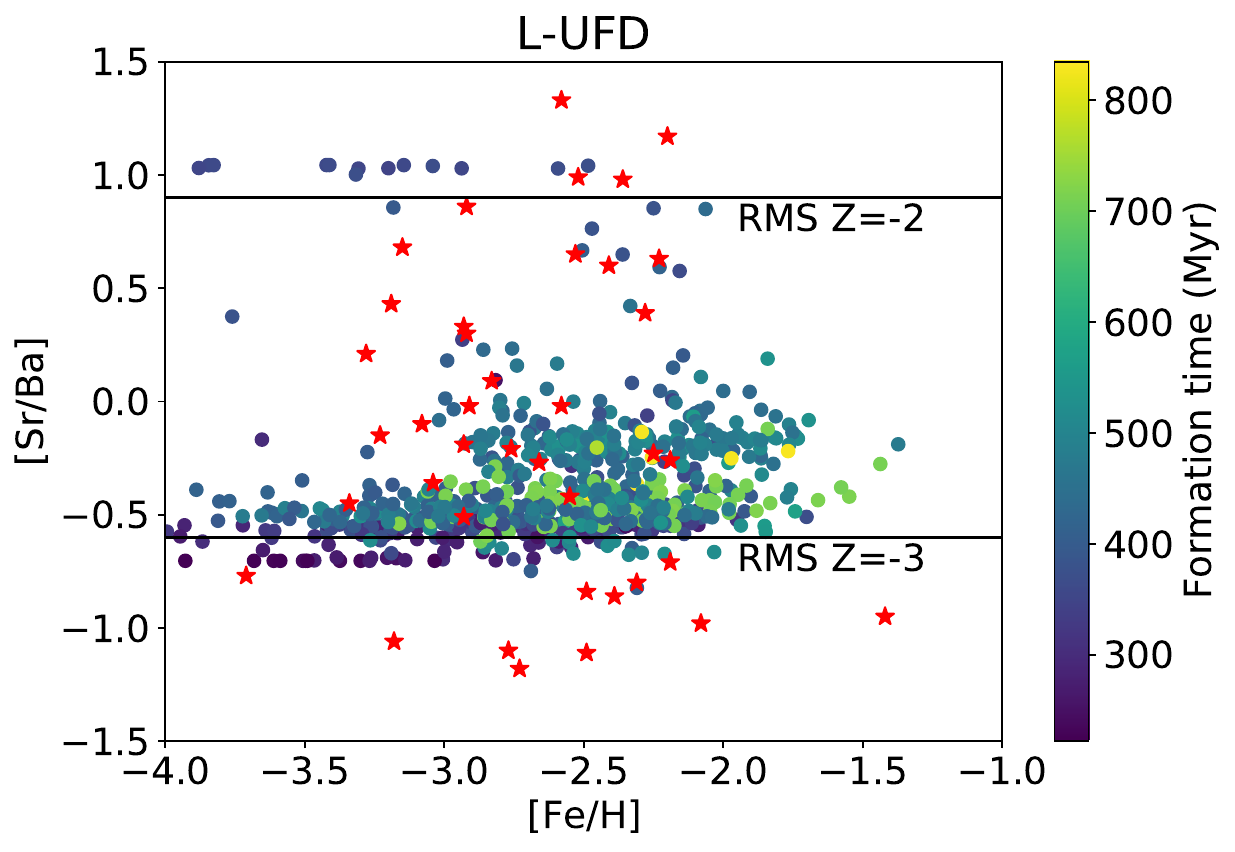}
    \includegraphics[width=\columnwidth]{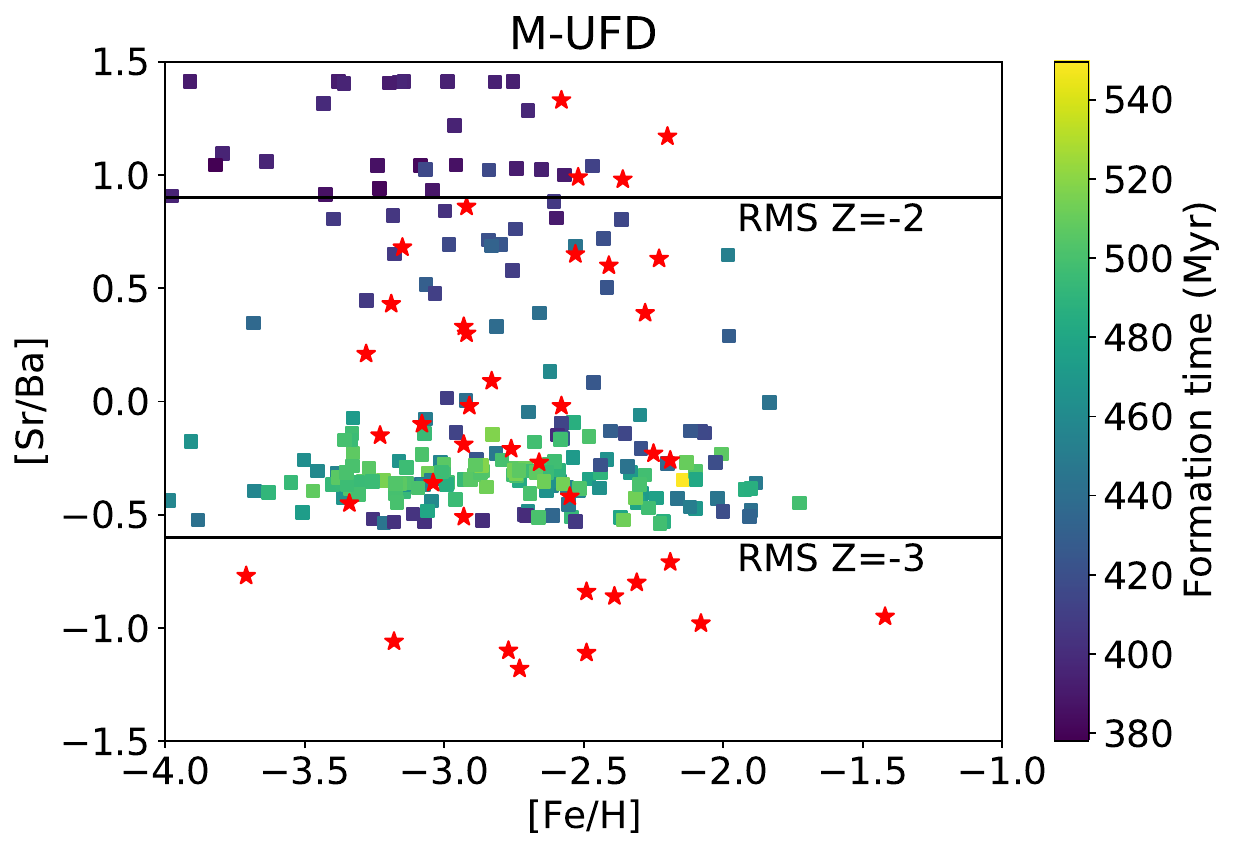}
    \caption{\revise{[Sr/Ba] ratios of simulated and observed stars, overlaid with  IMF-averaged values (horizontal lines). Accordingly, simulated [Sr/Ba] values for individual stars do not necessarily need to fit between these lines.}}
    \label{fig:SrBa ratio}
\end{figure*}

An interesting chemical feature of the RMS contribution is the evolution of the [Sr/Ba] ratio. In the \citet{2018Limongi_Chieffi_RMS} model, fast-rotating (300\,km\,s$^{-1}$) massive stars produce ejecta with $\mathrm{[Sr/Ba]} \sim -0.6$ while that of slow-rotating (150\,km\,s$^{-1}$) massive stars is $\mathrm{[Sr/Ba]} \sim 0.9$, for metallicities $-3<\mbox{[Fe/H]}<-2$. In Fig.~\ref{fig:SrBa ratio}, we show the simulated [Sr/Ba] ratios using the RMS model in the L- and M-UFDs. The median metallicities are $\mbox{[Fe/H]}=-2.66$ and $\mbox{[Fe/H]}=-2.96$, respectively. These [Sr/Ba] ratios are close to the IMF-averaged value at $\log(Z/Z_{\odot})=-3$.

\subsection{Adding electron-capture supernova yields}

\begin{figure}
    \centering
    \includegraphics[width=\columnwidth]{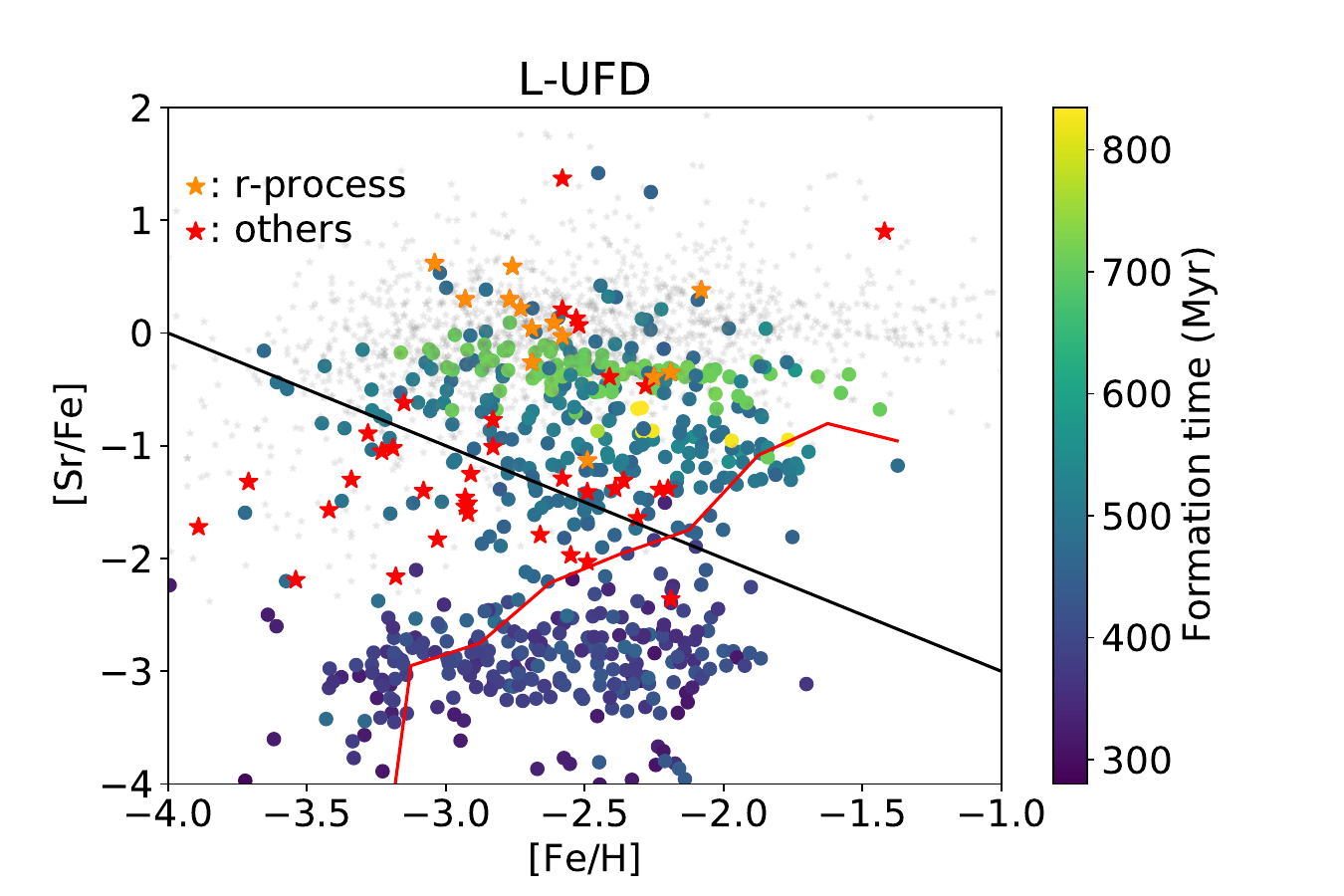}
    \caption{Observed abundances in UFDs overlaid with our ECSN model results for [Sr/Fe]. Symbols and colours are the same as in Fig. 2. The black line shows $\mbox{[Sr/H]} = -4$.} Sr production is quite efficient resulting in clearly separated abundance levels before and after the first ECSN explodes.
    \label{fig:ECSN}
\end{figure}

We test how our fiducial model results change when including an additional Sr source originating from ECSNe. Yields are taken from \citet{2018Wanajo_ECSNe}.
In our L- and M-UFDs, ECSNe occur five times and once, respectively. In Fig.~\ref{fig:ECSN}, we show the resulting Sr abundances.
The amount of Sr produced in one ECSN is quite large, therefore it triggers a strong enhancement in [Sr/Fe] in stars formed subsequently. For the L-UFD this leads to a significant gap at $[\mathrm{Sr}/\mathrm{Fe}] \sim -2$ in the simulated abundance distribution. Many of the observed abundances are roughly in that gap. The median of the simulated abundances steeply increases between $-3 < \mathrm{[Fe/H]} < -2$. However, this is a merely a results of the fraction of high [Sr/Fe] stars gradually increasing with [Fe/H].

In the M-UFD the ECSN occurred only at the very end of star formation, and thus did not affect the simulated Sr abundance. Later time stars would form from Sr enriched gas, though.

\section{Discussion}

\subsection{No {\it r}-process contribution to Ba and Sr}

The {\it s-}process enrichment in the MW  has been studied for decades (e.g. with chemical evolution models \citealt{2006Cescutti_MW_sprocess}). In our \revise{fiducial} model, the majority of
{\it s-}process elements is produced in low-mass stars with $M \lesssim 3\,\Msun$.
Because of the long delay-time and the strong metallicity-dependence of Ba yields from  AGB stars, other production paths such as that of the {\it r-}process are preferred as the origin of Ba in low-metallicity halo stars  with $-3 < $ [Fe/H] $ < -2$ (e.g. \citealt{2006Cescutti_MW_sprocess}). 
Indeed, the large scatter of Eu abundances among low-metallicity halo stars
can be explained by rare {\it r-}process enrichment \citep{2021Cowan_RprocessReview}. A similar conclusion is obtained by \citet{2019Hirai_Sr_in_dwarfs}, who study the production of Sr in classical dwarf galaxies. 

Among the stars found in UFDs,
there are no stars present with $\mbox{[Fe/H]} \gtrsim-1.5$, and many have $\mbox{[Fe/H]} \sim-3.0$. Considering the small stellar masses
of the UFDs, we infer that
the number of {\it r-}process events in a UFD is likely one at most. This implies that only a few systems actually experience an {\it r}-process event.
The stochasticity of events has been used to explain e.g. the \revise{very} high Eu abundance of stars in Ret~II, and an apparent `jump' in the Eu abundance observed in Gru~II. The moderate Eu abundances of Tuc~III stars could also be explained similarly by assuming that an r-process event occurred in the outskirt \citep{2020Tarumi}, or that the r-process element abundances were somewhat diluted by a larger mass of hydrogen than considered in the case of Ret~II \citep{Marshall18_TucIIIobservation}. 

All those stars are identified as r-process stars due to their strong Eu line (and consequently large [Eu/Fe] abundance), but they also have significant [Ba/Fe] because the [Ba/Eu] ratio of $-0.71$ is governed by the universal, nuclear physics based main r-process pattern \citep{2009Sneden}. The origin of strontium is less clear because a number of processes are 
thought to produce Sr,
but a similar conclusion can be drawn.

This leaves two implications, 1) Faint low-metallicity stars in UFDs with moderate S/N spectra may not result in a Eu detection. 
If the intrinsically stronger Ba line can be detected and a high [Ba/Fe] is obtained, then the {\it r-}process is likely the dominant source of Ba; 2) If such stars do not have a high [Ba/Fe] or if only an upper limit is placed, then we can exclude a likely contribution from a rare and prolific {\it r-}process event that should  caused both high [Eu/Fe] and [Ba/Fe].

\begin{figure}
    \centering
    \includegraphics[width=\columnwidth]{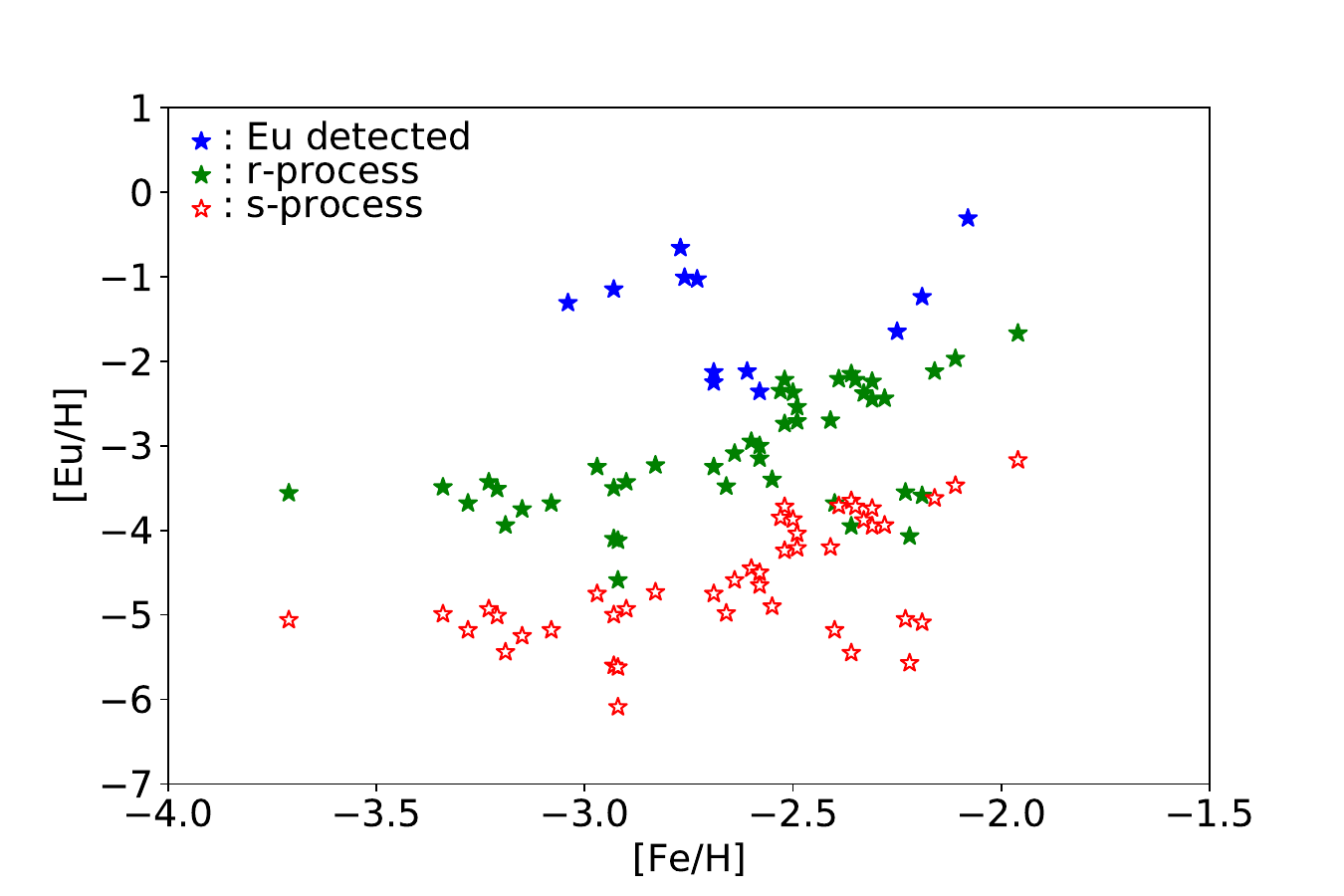}
    \caption{\revise{[Eu/H] distribution of various UFD stars as estimated from their [Ba/H]. Stars with actual Eu detections found in Ret~II, Tuc~III, and Gru~II are plotted with blue symbols. 
        For other UFD stars, each star produces two data points coloured green and red.
    The green and red symbols are plotted assuming $[\mathrm{Eu}/\mathrm{Ba}] = 0.7$ ({\it r}-process) and $-0.8$ ({\it s}-process), respectively.} }
    \label{fig:r-process}
\end{figure}

In Fig.~\ref{fig:r-process}, we show the expected [Eu/H] distribution of UFD stars. The blue symbols are stars with Eu detections. The green symbols are stars for which we assume that all their Ba originates with the main {\it r}-process. Their Eu abundances have been derived by assuming the {\it r}-process [Ba/Eu] = $-0.71$ ratio \citep{2008Sneden_review}. 
The red symbols depict the same stars but we now assume that they  have formed from gas enriched by the {\it s}-process. Here we assume [Ba/Eu] = $0.8$, which is calculated by the IMF-averaging of the yield of metallicity $\log(Z/Z_{\odot}) = -2.3$. We note that the values of some of the green symbols have higher [Eu/H] values than those of blue symbols. This suggests that if the origin of Ba in these stars were with the {\it r}-process, the Eu lines would have been detected and abundance measured. But this has not been the case. Indeed, for some stars the possibility of pure {\it r}-process has already been eliminated (e.g. Car~II, \citealt{2020Ji_CarII}). We thus make a working assumption and place an empirical detection limit for $[\mathrm{Eu}/\mathrm{H}] \sim -3.5$. The caveat is that this only roughly works for stars on the upper portion of the red giant branch, i.e. those stars sufficiently cool, which is satisfied here by these dwarf galaxy stars. For warmer stars with higher temperatures, detections of any intrinsically weak lines such as the Eu line at 4129\,{\AA} will be essentially impossible. This is also driven by stellar brightness as dwarf galaxy stars are generally very faint which precludes obtaining stellar high-resolution spectra of sufficient quality.
Nevertheless, in this sample considered here, and the fact that the Eu abundances obtained for stars from putative s-process material (red symbols) and those with actual Eu detection (blue symbols) are about one dex apart at least, clearly showcases why only few UFD stars have Eu measurements (owing to the significant Eu enrichment) when most do not.

\subsection{Production of {\it s}-process elements in the fiducial model}

To assess if there is any s-process contribution among low-metallicity stars in UFDs with low [Ba/Fe], we compare the stellar Ba abundances of the observed and simulated UFDs. 
For our simulated UFDs, we find that medians of [Ba/Fe] $\sim -2.5$ (L-UFD) and [Ba/Fe] $\sim -4$ (\revise{M-}UFD) for stars with $-3 < $ [Fe/H] $ < -2$, as can be seen in Fig.~\ref{fig:abundance scatter plot}. The values of [Ba/Fe] are the same as the yields from an SSP integrated from birth to an age of 100\,Myr (L-UFD) and \revise{25}\,Myr (M-UFD). This highlights that Ba can get contributed on short timescales.
We obtain similar results if we use [Sr/Fe] instead. Note that these
estimated ages are insensitive to the exact choice of the yields, as it does not affect the star formation in the galaxy.
If we were to change the Ba yields of AGB stars, we would expect that the typical [Ba/Fe] of stars formed in these galaxies are again
represented by the [Ba/Fe] level produced by an SSP at 100 (25)\,Myr in the L- (M-) UFDs as chemical enrichment does not have much of an effect on the star formation history.
\revise{Therefore, we argue that an additional Ba source contributes to the chemical enrichment with an approximate, average delay of $\sim 100\,\mathrm{Myr}$.}

The duration of star formation in UFDs can be constrained to up to a few hundred\,Myr because gas heating by cosmic reionization can effectively quench star formation in these small galaxies by $z \sim 6$. In addition, considering their low halo masses, we do not expect the onset of star formation in UFDs to be very early. 
In our simulations, star formation typically begins when the cosmic age is about 300\,Myr. With these conditions the duration of star formation in UFDs is shorter than 700\,Myr. In our simulations, the L-UFD has a comparable length of star formation. Therefore the L-UFD can be considered as one of the most {\it s-}process enhanced UFDs. However, the predicted Ba abundance \revise{in the fiducial model} is still lower than the observed values (red and orange stars; see Fig.~\ref{fig:abundance scatter plot}). This suggests that Ba needs to be produced more efficiently in the low-metallicity environments present in UFDs.

A simple resolution to enhance Ba production might be to adopt different AGB yield models.
It may solve the inconsistency if the yields were $10 \hyphen 100$ times higher. However, such large amounts are unlikely 
because the yield of {\it s}-process elements in AGB stars is roughly proportional to the abundance of seed nuclei at low-metallicity.
Such enhanced AGB yields would also lead to an overproduction of Sr, and also of Ba at higher metallicities ($-2 < \mbox{[Fe/H]}$).

\begin{figure}
    \centering
    \includegraphics[width=\columnwidth]{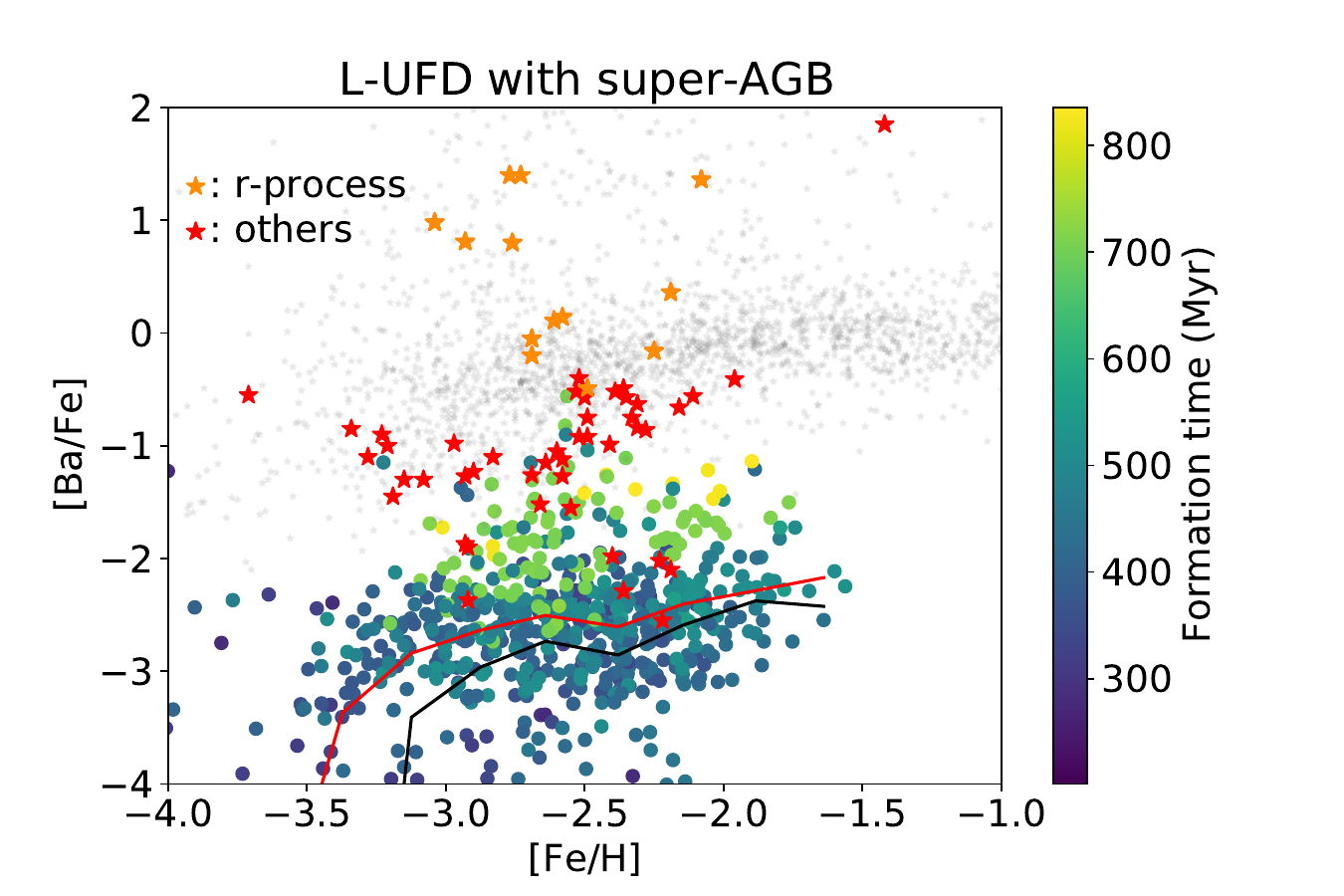}
    \includegraphics[width=\columnwidth]{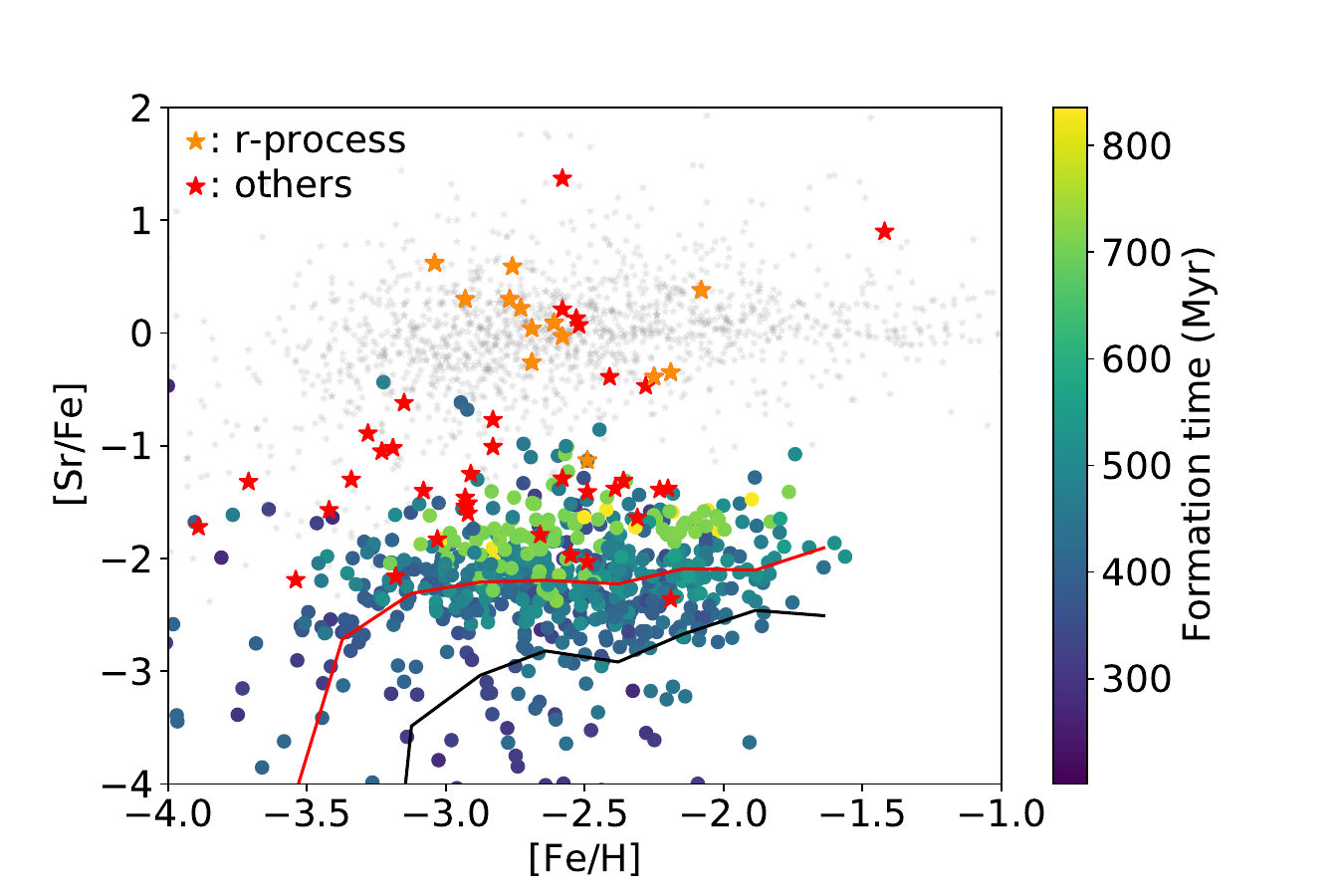}
    \caption{Results based on a model that includes Ba \revise{and Sr} production from super-AGB stars in addition to the one from AGB stars. The red and black curves show the model result with and without any contribution from super-AGB stars. The super-AGB stars
    enhance [Ba/Fe] \revise{and [Sr/Fe]} slightly but not sufficiently to explain the observed UFD abundances. }
    \label{fig:sAGB}
\end{figure}

An alternative candidate for a Ba source might be super-AGB stars \citep{2017Doherty_superAGB}. The typical progenitor mass is around $6 \hyphen 8$\,\Msun.  Thus, super-AGB stars produce {\it s-}process elements within 100\,Myr of their birth. 
We employ the {\it s-}process yields from \citet{2017Doherty_superAGB} (C. Doherty 2020, priv. comm.)
and re-calculate our simulated stellar elemental abundances. 
We use the yield of super-AGB stars at metallicity of $\mbox{[Fe/H]}=-0.7$. 
We find only a minor contribution of super-AGB stars to the production of Sr,
as shown in Fig.~\ref{fig:sAGB}. This is owing to the narrow mass range of the progenitors and a less efficient production of {\it s-}process elements in super-AGB stars.
Although super-AGB stars enable early 
{\it s-}process enrichment, the absolute amounts of Ba and Sr in UFDs do not increase significantly.
In order for super-AGB stars to be the main source of {\it s}-process elements in UFDs, the yield at [Fe/H]$\,\sim -2$ would need to be enhanced by a factor of $\gtrsim$ 10 compared with the one at [Fe/H]$\,= -0.7$.

Sr is lighter than Ba and can thus be synthesised in less extreme conditions (e.g. in a `light-element primary process', \citealt{2007Montes_LEPP}). Such a source for Sr 
may be necessary to explain the existence of stars with high [Sr/Ba] and [Sr/Eu] ratios \citep{2004Honda_hondastar}.
\cite{2019Hirai_Sr_in_dwarfs} consider ECSNe and RMSs as additional sources of Sr. They conclude that the additional Sr production 
in short-lived stars is necessary to explain the Sr abundances found in dwarf galaxies. 

In our \revise{fiducial model}, Sr is only produced in AGB stars. Fig.~\ref{fig:abundance scatter plot} shows that Sr behaves similarly as Ba.
Sufficient amounts of Sr are not produced in our models. 
To explain the observed Sr abundances ([Sr/Fe] $\sim$ $-1$), 
we can estimate the required Sr yield of $10^{-8}\,\Msun$\ per 1\,\Msun\ of stars in the first 100\,Myr. But considering the large differences of Sr abundances among different UFDs, we speculate that besides the r-process there may well exist an additional production source of Sr that is rare and prolific.

\subsection{Exploring additional Ba and Sr sources -- RMS}

\revise{RMSs can be the origin of both Ba and Sr, produced through the weak and main {\it s}-process. The rotation speed determines the ratio between Ba and Sr. }
\revise{In the L-UFD, we can see an increasing trend in the [Sr/Ba] ratio. This is a consequence of the contribution from the component with $\log(Z/Z_{\odot})=-2$ playing a role. The smaller M-UFD seem to show an inverse trend, and it is a stochastic effect due to the incomplete sampling of IMF and IDROV. Stars formed later shows [Sr/Ba] ratio converged to the IMF-average value at $\log(Z/Z_{\odot})=-3$.}

\revise{At $\mbox{[Fe/H]} \gtrsim -2.5$, [Sr/Ba] increases to $-0.2$.
This trend reflects the fact that at $\log(Z/Z_{\odot})=-2$, the fraction of stars rotating with 300\,km\,s$^{-1}$ has greatly decreased compared to  $\log(Z/Z_{\odot})=-3$. On the contrary, the fraction of stars rotating with 150\,km\,s$^{-1}$ does not change significantly, rendering the overall contribution of slower rotation stars dominant.}

\revise{We can infer the fraction of fast rotators at $-3 \lesssim \mathrm{[Fe/H]} \lesssim -2$ from simulated Ba abundances, as it is synthesised primarily in fast rotators. The fraction should be $\sim 10$ per cent to explain the observed Ba abundances in UFDs. If the fraction is decreased or increased by 1\,dex (1 or 100 per cent), Ba is under- or over-produced and thus incompatible with the observations. Since this constraint comes from the IMF-averaged amount of produced Ba, it is degenerate with the yields of RMS. The Ba abundances in UFD stars favour $\sim 1\times 10^{-9}$\,\Msun of Ba production per 1\,\Msun of stars formed. We obtain a similar constraint on Sr, and the IMF-averaged Sr production should be $\sim 1\times 10^{-8}$\,\Msun per 1\,\Msun of stars formed.}

\revise{Although RMS roughly explain the [Ba/Fe] and [Sr/Fe] trends of UFDs as a whole, there remains some tension between the simulated and observed [Sr/Ba] distribution. The observed [Sr/Ba] values appear to be independent of [Fe/H]. However, if RMS are the dominant {\it s}-process source, we expect an increase of [Sr/Ba] as [Fe/H] following the slow-down of the rotation. Although the stellar masses of UFDs are too small to well-sample the IMF, all UFDs as a whole have sufficient masses, and we could detect the metallicity trend if it exists.}

\revise{Another important test can be a comparison with observed nitrogen abundances. There are a recent observations by \citet{2020Ji_CarII} who use a CN molecular feature to obtain N abundances for three stars in Car~II. They find no correlation of N with Ba or Sr. If RMS were the dominant producers of {\it s}-process elements, then stars enhanced in {\it s}-process elements should be also enhanced in N because N is readily produced in copious amounts due to fast rotation \citep{2006Meynet_rotation_CNOcycle}. The stellar mass of Car~II is $\sim 10^{4}\,\Msun$ (similar to our L-UFD), and therefore it is massive enough to average-out the effect of RMS to its IMF-averaged value. Although the  N measurement uncertainties are quite large ($\gtrsim 0.5$\,dex) and the number of current observations remains limited, it is quite interesting to determine abundances of N in UFD stars, as an ongoing test for the RMS scenario. UFDs are particularly important because if RMSs exist, they would dominate the production of slow neutron-capture elements and N.}

\subsection{Exploring additional Sr sources -- ECSNe}

\revise{ECSNe produce a copious amount of Sr. As we have shown in Fig.~\ref{fig:ECSN}, the amount is so large that even an ECSN imprints a simple and conspicuous signature. The ECSN progenitor mass-range is unconstrained, and it is quite difficult to determine the mass-range from theoretical stellar evolution modelling. }

\begin{figure}
    \centering
    \includegraphics[width=\columnwidth]{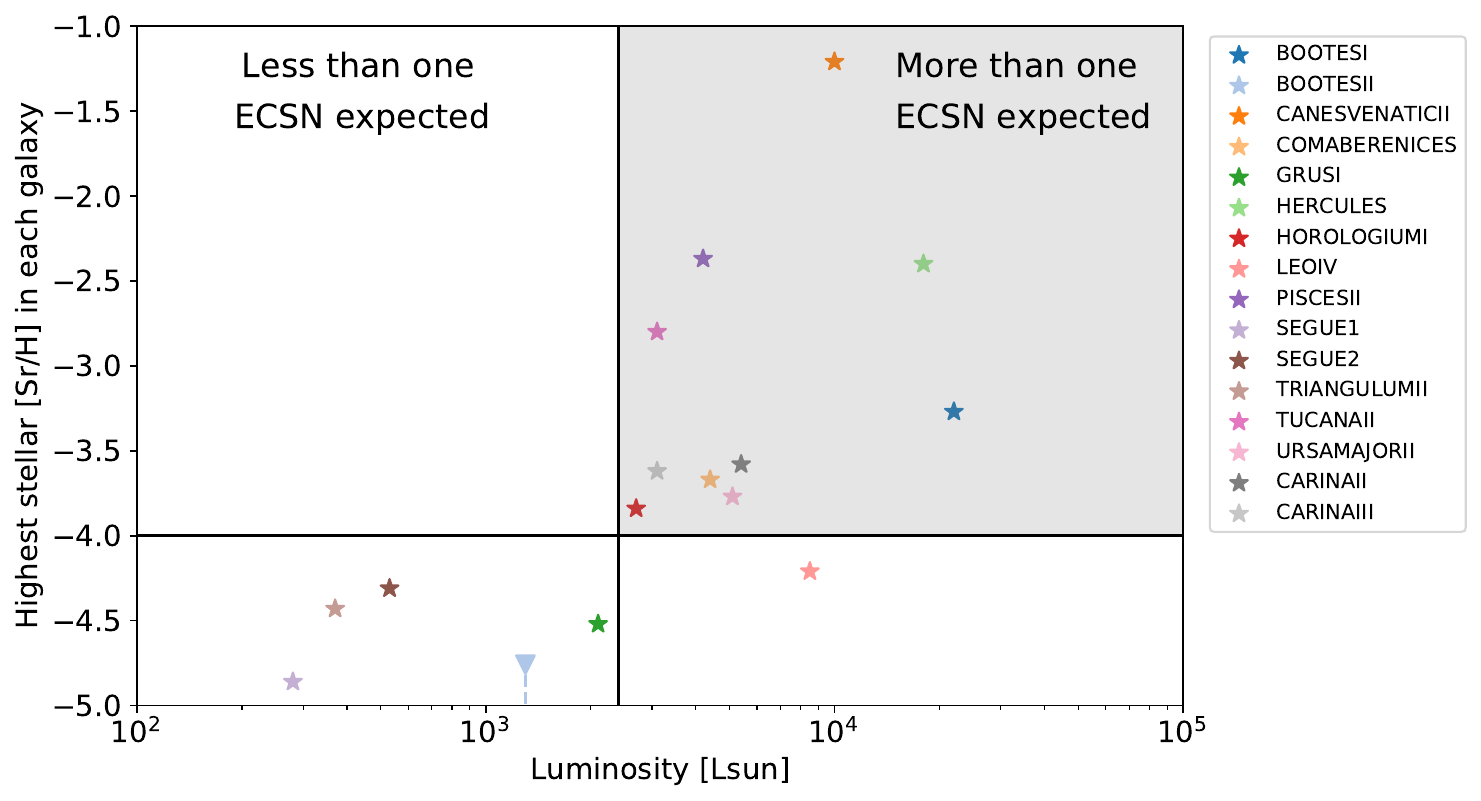}
    \caption{The highest value of stellar [Sr/H] in each UFD against luminosity.
    Various symbols and stars represent Sr abundances. We show the stars with highest abundance observed in each UFD.
    The downward triangle depicts the highest upper limit because no Sr detection have been reported for this UFD.
    One CH star in Segue 1 has been removed from the analysis as it is reflective of the abundances of the binary companion.}
    \label{fig:Sr_luminosity}
\end{figure}

\revise{ECSNe are rare and prolific sources of Sr. Assuming that the ejected Sr is diluted with less than $10^{7}\,\Msun$ of metal-free gas, a single ECSN would enrich a galaxy to $\mathrm{[Sr/H]} \gtrsim -4$. Without any contribution from ECSNe, low-mass UFDs would have very low Sr abundances ($\mathrm{[Sr/H]} \lesssim -4.0$).  Segue I, Segue II, Triangulum II, Bootes II, Grus I are considered to be such UFDs. Tucana II is also such a UFD if the Sr-rich star (Tuc-033) is to be regarded as a foreground contaminant. 
The observed [Sr/H] and luminosities of UFDs clearly show such trend (Fig.~\ref{fig:Sr_luminosity}). The five faintest UFDs have $\mathrm{[Sr/H]} < -4$. We argue that the very low Sr abundances in the faintest UFDs are the consequence of no ECSNe having occurred in these systems. More massive UFDs could have experienced one or more ECSN as indicated in Fig.~\ref{fig:Sr_luminosity}}.

\revise{We can place an upper limit to the ECSN rate based on the Sr abundances of UFD stars. From Fig.~\ref{fig:Sr_luminosity}, we infer that galaxies with $\lesssim 2400\,\mathrm{L}_{\odot}$ should not have hosted any ECSNe. Then we can constrain the ECSN rate to be roughly less than one per $2000\,\mathrm{L}_{\odot}$ of stars. Assuming that the stellar mass to luminosity ratio is roughly unity, the current surviving stellar masses of UFDs are $\sim 2000\,\Msun$. 
Although UFDs generally could have mass-to- luminosity ratios larger than unity, we here
derive a rough constraint.
Assuming that 40 per cent of stars formed have survived until the present day (assuming a Chabrier IMF), the original stellar masses should have been 5000\,\Msun. Therefore, the ECSN rate can be constrained to be less than 1 per 5000\,\Msun of stars formed. It roughly corresponds to the progenitor mass range of $\Delta M$ of 0.1\,\Msun. For example, if only stars with 8.0 -- 8.1$\,\Msun$ explode as ECSNe, the rate would be 1/5000\,\Msun. }

\revise{Interestingly, some UFDs are highly Sr-enhanced: Pisces II, Hercules, and Canes Venatici II. ECSNe could be the origin of such prominent Sr enhancement. In our simulations, 
we find several stars with $\mathrm{[Sr/Fe]} \sim 0$ are formed in the L-UFD. Because of the lower energy of ECSN ($\sim 10^{50} \mathrm{erg}$), the minimum dilution mass is likely small, e.g. $\sim 10^{4}\,\Msun$. In this case, stars could be formed from gas enriched to $\mathrm{[Sr/H]} \sim -1$ with a single prior ECSN, explaining the high [Sr/Fe] reported by \citet{2016Francois_BaFe}.}

\subsection{IMF variation}

The IMF of stars in metal-poor environments such as UFDs could be different from the one governing metal-rich environments \citep{2013Kroupa_IGIMF_book, 2018_shallow_IMF_in_UFD}. \citet{2009Komiya} propose a log-normal IMF centred at 3 -- 20\Msun to explain the fraction of {\it s}-process enhanced stars among carbon-enhanced metal-poor stars in the MW. \revise{\citet{2013Geha_shallowIMF} argue that a shallower IMF slope can explain the colour-magnitude diagrams of UFDs. Theoretically, the IMF of metal-free stars is expected to be top-heavy \citep{2014Hirano_100firststars}. The stellar IMF might have been top-heavy in low-metallicity galaxies like UFDs. In contrast, \citet{2012Conroy} suggest that a bottom-heavy IMF would explain the Na$_{I}$ (0.82 $\mu\,$m), Ca$_{\rm II}$ (0.86 $\mu\,$m), and FeH (0.99 $\mu\,$m) spectral features observed in early-type galaxies}.

\begin{figure}
    \centering
    \includegraphics[width=\columnwidth]{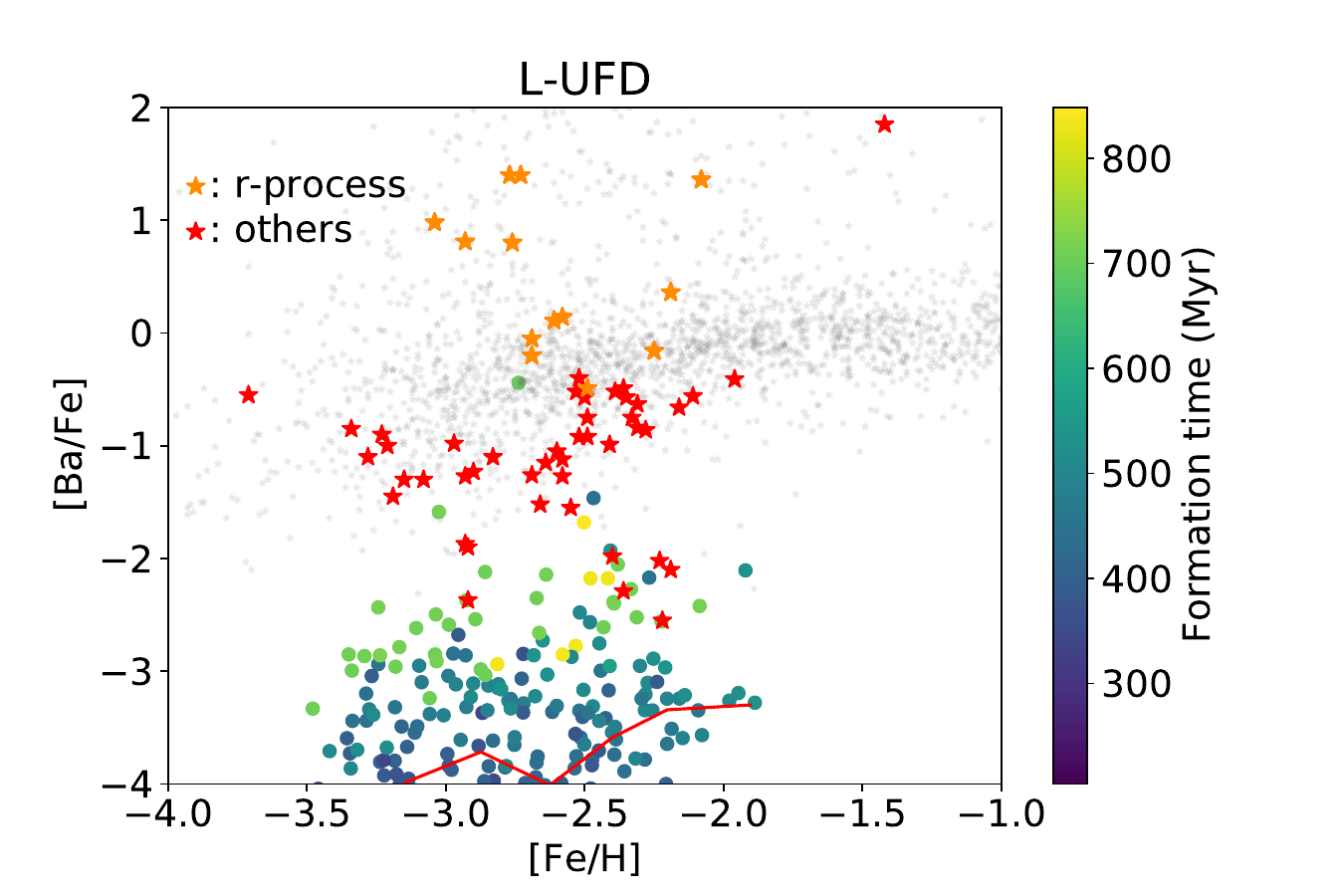}
    \includegraphics[width=\columnwidth]{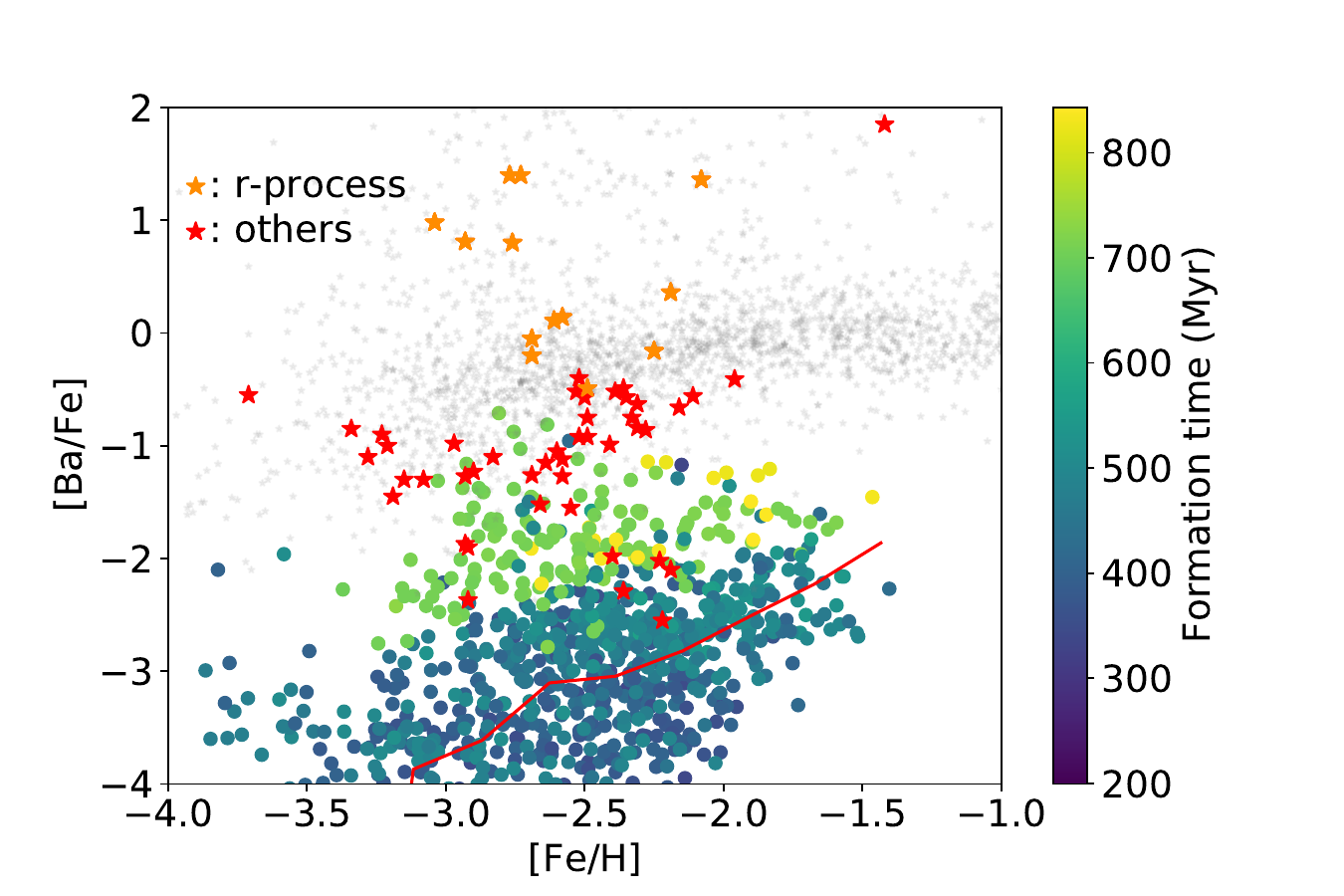}
    \includegraphics[width=\columnwidth]{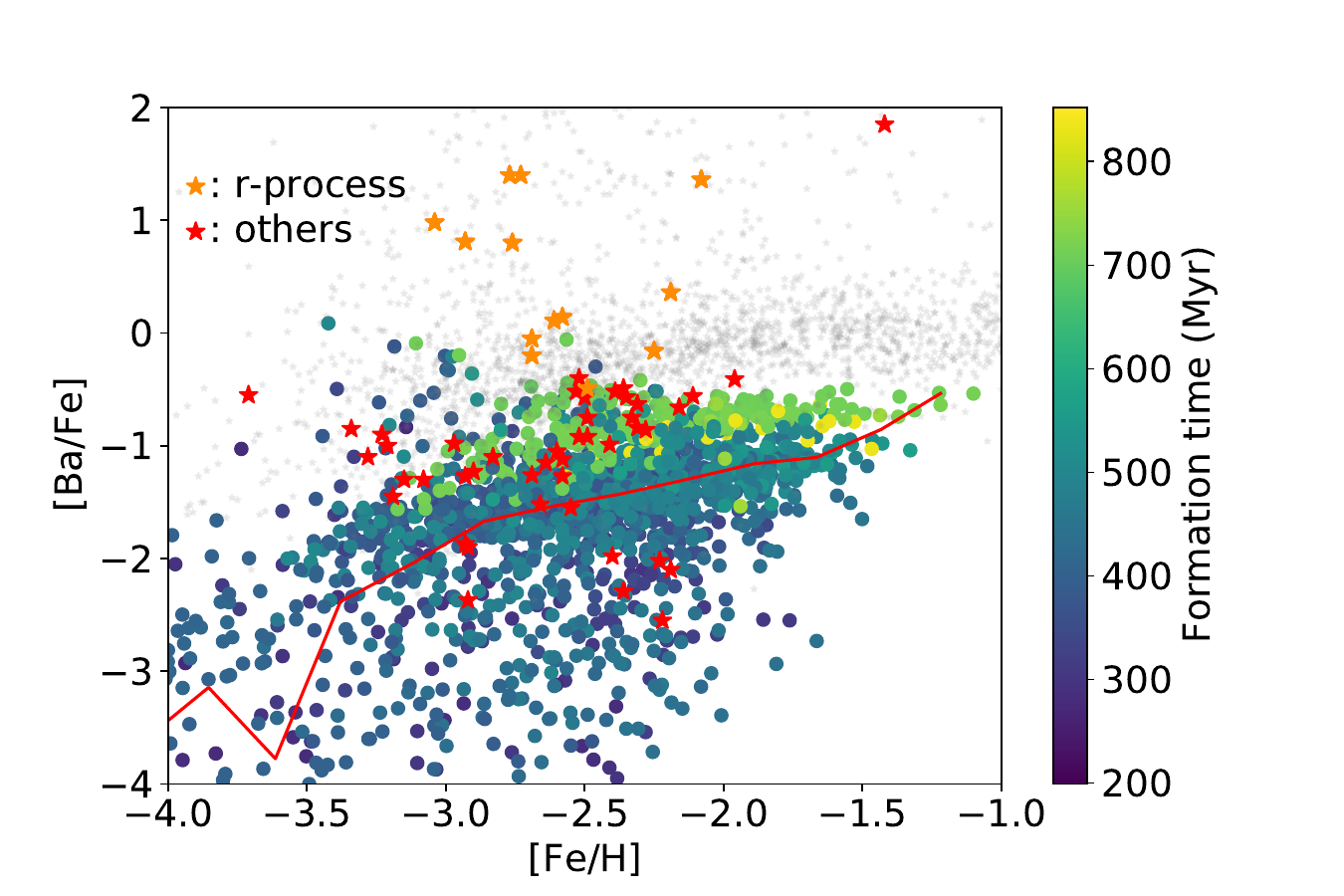}
    \caption{\revise{Results when testing the effects of varying the IMF. Top panel: top-heavy (log-flat) IMF model. Ba production per stellar mass slightly increases but Fe production increases by a larger degree, thus lowering the [Ba/Fe]. Middle panel: bottom-heavy (Salpeter) IMF. The difference to a Chabrier IMF is only among the low-mass ($\lesssim 1 \Msun$)    stars that do not contribute to the chemical enrichment of the galaxy regardless. Prediction thus resemble that of our fiducial run. Bottom panel: Komiya-like IMF (log-normal, centred at 4\,\Msun and a standard deviation of 0.15\,dex). Ba production is enhanced and Fe production is suppressed at the same time. Simulated [Ba/Fe] values roughly match the ones of UFDs.} Observational data are shown with the same symbols as in Fig.~\ref{fig:abundance scatter plot}.}
    \label{fig:modified_IMF}
\end{figure}

As a simple test \revise{for possible IMF variations, we run three additional simulations with different IMFs: top-heavy (log-flat), Komiya-like (log-normal, centred at 4\,\Msun\ and the standard deviation of 0.15\,dex), and bottom-heavy \citep{1955_Salpeter}. In Fig.~\ref{fig:modified_IMF}, we show the results of these tests. The [Ba/Fe] values for the model with the top-heavy IMF are lower than those for our fiducial model (see Fig.~\ref{fig:abundance scatter plot}), because 
the Fe production is enhanced, whereas the Ba production remains similar to that with the Chabrier IMF. Note that the top-heavy IMF model results in a similar distribution of [Fe/H] compared with the fiducial model in spite of its more efficient iron production. This is because the stellar mass formed in the UFD decreases by a factor of 2 to 3 compared with the fiducial model as a result of the enhanced stellar feedback, but the total amount of Fe remains similar. 
In the case of the bottom-heavy IMF, the effective mass distribution is almost the same at around the initial mass $M$ $\gtrsim 1\,\Msun$ which is the mass range that contributes most to Ba enrichment.
Therefore the result is almost the same as that of our fiducial model. 
Using a Komiya-like IMF increases the number ratio of intermediate-mass stars to massive stars, and accordingly, supplies more barium and less iron. This reproduces the observed trend of [Ba/Fe] better as can be seen in Fig.~\ref{fig:abundance scatter plot}.}
We note that, in addition to the stars represented by this IMF, low-mass companions in binary systems are assumed to be formed and to survive until today, although the companions are not explicitly represented in our simulations.

\revise{Integrated-galaxy IMF theory provides us with a systematic way to discuss IMF variations from SFR \citep{2013Kroupa_IGIMF_book}. In vigorously star-forming clouds, high-mass stars preferentially form, whereas slowly star-forming clouds produce a large number of low-mass stars. The star formation duration in our UFD simulations is $\sim 100$\,Myr, and the total stellar mass is $\sim 10^{4}\,\Msun$, therefore the star formation rate is $10^{-4}\,\Msun$ yr$^{-1}$. This SFR is low, and a suppression of the formation of stars with $\gtrsim 10\,\Msun$ is expected (see Fig.~35 of \citealt{2013Kroupa_IGIMF_book}). This suppresses the production of Fe while holding Ba and Sr steady, therefore an IMF similar to our best-fit IMF might be a natural consequence of low SFRs in these small systems. An important test is to apply the same framework of the IMF variations to other systems with different SFRs. We will further investigate this in our future study.}

Regarding the peculiar IMF we have considered, there may be an interesting connection to the multiple populations in globular clusters (GCs).
Their chemical abundances could be explained by contributions from super-AGB stars
to the overall chemical enrichment before the formation of any second-generation stars (e.g. \citealt{2018Bastian}). There are also implications for other issues
such as the mass-budget problem, in which the ratios of the total mass of first-generation stars to second-generation stars is anomalous 
\citep{2008Renzini_MassBudgetProblem}.
The log-normal IMF, which has a large mass fraction of super-AGB stars, could alleviate the mass-budget problem. Also, the log-normal IMF greatly reduces the production of Fe. Therefore, the metallicities do not significantly differ between the first-generation and the second-generation stars. 
A caveat to concluding that the IMF presented by \citet{2007Komiya} is a viable solution for GCs would be that the the majority of GCs stars have [Fe/H] $> -2$ which is outside of the range for which this IMF has been developed (\citealt{2013Suda_IMF, 2014Lee_IMF}).
Nevertheless, it is interesting that enhancing the fraction of super-AGB stars while decreasing the number of massive stars is also favoured in a completely different context of chemical evolution. 

Considering the observed Ba abundances of these UFD stars, an additional source of Ba is required that is not operating an {\it s}-process and not a prolific {\it r}-process. This leaves room for
another {\it r}-process source with lower yields.
Recent observations reveal that [Ba/Fe] of MW stars does not increase with [Fe/H] in the range of $-2.0 < $ [Fe/H] $ < -1.5$, suggesting that {\it s-}process yields from AGB stars do not contribute to chemical evolution for stars formed with [Fe/H] $ < -1.5$ \citep{2020Matsuno_no_sprocess}. 
However, the [Ba/Eu] of these sample stars is higher than the value predicted by the main {\it r-}process nucleosynthesis. This may suggest some {\it s-}process events which produce Ba with a delay-time similar to the timescale of the metal enrichment leading to [Fe/H] $\sim -1.5$. An interesting test of this possibility would be to measure or place strong constraints on Eu abundances for stars with measurements of Ba abundances in UFDs, so that we can identify the main contributors of Ba as originating from either the {\it r-} or {\it s-}process(es).

\subsection{On UFDs with internal Ba spreads}

Boo~I, Car~II, and ComBer show clear internal spreads in [Ba/Fe]. The reason for the Ba spreads could be either temporal- or spatial inhomogeneities. If it is temporal (see Fig.~\ref{fig:abundance scatter plot}), AGB stars might be the source, and then star formation in these systems is likely a prolonged one ($\gtrsim 300\ \mathrm{Myr}$). 
Fig.~\ref{fig:abundance scatter plot} shows that AGB stars can indeed cause an internal spread in a galaxy. There is a clear difference in the resulting [Ba/Fe]. In our simulations, even the stars formed later have $\mathrm{[Ba/Fe]} \sim -1.5$, but some stars in Car~II or Boo~I have $\mathrm{[Ba/Fe]} \sim -0.5$. The large difference may indicate that the star formation in these systems is even more prolonged ($\gtrsim 1\ \mathrm{Gyr}$) than what is realised in our simulations. 

In the case of spatial inhomogeneities (see Fig.~\ref{fig:Rotating massive stars}), fast-rotating massive stars could be the origin of Ba in these systems. They produce sufficient amounts of Ba and produce some stars with $\mathrm{[Ba/Fe]} \sim -0.5$. The star formation duration should then be relatively short ($\lesssim 200\ \mathrm{Myr}$). However, in this model, [Ba/Fe] does not show an increasing trend with [Fe/H], and therefore, cannot reproduce what is observed for Boo~I. A possible resolution might be to assume a different distribution of stellar rotation speeds. A higher fraction of fast-rotating stars at $\log (Z/Z_\odot)=-2$ could reproduce the increasing trend of [Ba/Fe] in Boo~I. An important test of this non-standard scenario would be an quantifiable correlation of nitrogen abundance with Sr or Ba, as it is necessary to observationally confirm the co-production of these elements. Accurate measurements of nitrogen abundances would allow us to answer the question.

\subsection{Effects of tidal stripping}

Signatures of tidal stripping are observed in some UFDs. If tidal stripping is common among UFDs, their stellar masses could potentially be significantly higher than what we assume ($\sim 10^{5}\,\Msun$) before the stripping. In an extreme case, they could have been so massive that the stochasticity of {\it r}-process would no longer apply, and the enrichment events would have become averaged out. In addition, their gravitational potential wells could have been significantly deeper, allowing them to keep gas and forming stars after cosmic reionization. Therefore, our conclusions regarding short star formation timescales and the `0 or 1'-type behaviour of {\it r-}process enrichment would not hold. Boo~I and Ursa Major~II, which have stars with [Ba/Fe] $ > -1$, may indeed have lost stellar mass due to tidal forces by the MW \citep{2006_Boo1, 2006_UMa2}. 

In the scenario of initially more massive UFDs, they must have been able to form stars even after cosmic reionisation. Such galaxies accumulate {\it s}-process elements. Also they would have had enough mass to average out the stochastic nature of {\it r-}process events. Accordingly, the shortcoming of our models to reproduce stars with [Ba/Fe] $\sim -0.5$ might possibly imply that Ba-rich UFDs are actually such tidally stripped systems. However, most UFDs are too far away from the Galactic centre to lose their stars in this way \citep{2018Simon_tidalstripping}. Also, there are UFDs with [Ba/Fe] $\sim -1$ that do not show signatures of tidal disruption or past stripping events. Hence, in the present paper, we assume that the UFDs have retained most of their mass over time, including their early stellar populations.

\section{Conclusions}

We have modelled the evolution and chemical enrichment process of small, early galaxies using moving-mesh hydrodynamic simulations. We have focused on the production of Ba in UFDs, but we have also considered origins of Sr in order to compare our results with observed abundance patterns of stars in UFDs. 
Understanding the enrichment history or Ba and Sr is important because even the most up-to-date sets of elemental yields that can reproduce the {\it s-}process material in the MW, they cannot explain observed trends of Ba and Sr abundances in UFDs, likely owing to a very different star-formation and chemical enrichment history.

Our main findings are summarised as follows: 
\begin{enumerate}
    \item Ba production rate should be efficient with $\sim 1\times 10^{-9}$\,\Msun\ per 1\,\Msun\ of stars in the first 100\,Myr,  even in low-metallicity environments such as in UFDs. The required short delay-time suggests that the origin of Ba must have a short evolutionary timescale, which is fulfilled by short-lived, massive stars. Super-AGB stars alone cannot resolve the inconsistency between our simulation results and observations.
    \item \revise{Rotating massive stars (RMSs) can explain the Ba abundances of observed UFDs if $\sim 10$ per cent of stars are rotating at $\sim 300$\,km\,s$^{-1}$. An interesting test of such a RMS model would be provided by nitrogen abundance. If RMSs are the dominant factory of {\it s}-process elements, we would expect correlations between the abundances of nitrogen and {\it s}-process elements.}
    \item \revise{Electron-capture supernovae (ECSNe) are an efficient Sr factory. ECSNe leave a clear signature in the abundance patterns once occurred in small systems like UFDs. From the fact that the Sr abundances in low-luminosity UFDs are particularly low, we argue that the UFDs were not affected by ECSNe. Considering the UFDs' luminosities and stellar masses, the rate of ECSNe would be less than 2 per cent of CCSNe, narrowing the mass range of ECSNe progenitors
    to $\lesssim$ 0.1\,\Msun at $-3 < \mathrm{[Fe/H]} < -2$.}
    \item It is possible to explain the observed Ba abundances by introducing an IMF that produces predominantly super-AGB and AGB stars while suppressing the Fe production by massive stars. \revise{Such a top-light IMF might be realised in small galaxies with low star-formation rates \citep{2013Kroupa_IGIMF_book}}. 
\end{enumerate}
UFDs serve as an important laboratory for studies on nucleosynthesis in the early Universe. Future spectroscopic observations of additional stars in many UFDs, as well as measurements of other elemental abundances such as Eu and N, would allow us to identify the origin of Ba and other neutron-capture elements.

\section*{Acknowledgements}

We thank Robert Grand for his contribution to the earlier version of the manuscript and also for allowing us to use the Auriga galaxy formation model. We thank Volker Springel for kindly providing the simulation code \textsc{arepo}, Jill Naiman for assembling yields table for {\it s}-process elements, and Carolyn Doherty for kindly sharing her yields table of super-AGB stars.
We thank Shinya Wanajo, Yutaka Hirai, Kenta Hotokezaka, Yutaka Komiya, and  Masayuki Fujimoto for fruitful discussions. This study was supported
by World Premier International Research Center Initiative (WPI), MEXT, Japan and by SPPEXA through JST CREST JPMHCR1414. The numerical computations presented in this paper were carried out on Cray XC50 at Center for Computational Astrophysics, National Astronomical Observatory of Japan. TS was supported by JSPS KAKENHI Grant Numbers 20HP8012 and 16H02166. FvdV acknowledges support from the Deutsche Forschungsgemeinschaft through project SP 709/5-1 and from a Royal Society University Research Fellowship. AF acknowledges support from NSF grant AST-1716251, and thanks the Wissenschaftskolleg zu Berlin for their generous hospitality.

%%%%%%%%%%%%%%%%%%%%%%%%%%%%%%%%%%%%%%%%%%%%%%%%%%
\section*{Data Availability}

The data underlying this article will be shared upon reasonable request to the corresponding author.
%The inclusion of a Data Availability Statement is a requirement for articles published in MNRAS. Data Availability Statements provide a standardised format for readers to understand the availability of data underlying the research results described in the article. The statement may refer to original data generated in the course of the study or to third-party data analysed in the article. The statement should describe and provide means of access, where possible, by linking to the data or providing the required accession numbers for the relevant databases or DOIs.



\bibliographystyle{mnras}
\bibliography{sprocess} % if your bibtex file is called example.bib

\begin{thebibliography}{}
\makeatletter
\relax
\def\mn@urlcharsother{\let\do\@makeother \do\$\do\&\do\#\do\^\do\_\do\%\do\~}
\def\mn@doi{\begingroup\mn@urlcharsother \@ifnextchar [ {\mn@doi@}
  {\mn@doi@[]}}
\def\mn@doi@[#1]#2{\def\@tempa{#1}\ifx\@tempa\@empty \href
  {http://dx.doi.org/#2} {doi:#2}\else \href {http://dx.doi.org/#2} {#1}\fi
  \endgroup}
\def\mn@eprint#1#2{\mn@eprint@#1:#2::\@nil}
\def\mn@eprint@arXiv#1{\href {http://arxiv.org/abs/#1} {{\tt arXiv:#1}}}
\def\mn@eprint@dblp#1{\href {http://dblp.uni-trier.de/rec/bibtex/#1.xml}
  {dblp:#1}}
\def\mn@eprint@#1:#2:#3:#4\@nil{\def\@tempa {#1}\def\@tempb {#2}\def\@tempc
  {#3}\ifx \@tempc \@empty \let \@tempc \@tempb \let \@tempb \@tempa \fi \ifx
  \@tempb \@empty \def\@tempb {arXiv}\fi \@ifundefined
  {mn@eprint@\@tempb}{\@tempb:\@tempc}{\expandafter \expandafter \csname
  mn@eprint@\@tempb\endcsname \expandafter{\@tempc}}}

\bibitem[\protect\citeauthoryear{{Ad{\'e}n} et~al.,}{{Ad{\'e}n}
  et~al.}{2009}]{2009Aden_BaFe}
{Ad{\'e}n} D.,  et~al., 2009, \mn@doi [\aap] {10.1051/0004-6361/200912718},
  \href {https://ui.adsabs.harvard.edu/abs/2009A&A...506.1147A} {506, 1147}

\bibitem[\protect\citeauthoryear{{Ad{\'e}n}, {Eriksson}, {Feltzing}, {Grebel},
  {Koch}  \& {Wilkinson}}{{Ad{\'e}n} et~al.}{2011}]{2011Aden_BaFe}
{Ad{\'e}n} D.,  {Eriksson} K.,  {Feltzing} S.,  {Grebel} E.~K.,  {Koch} A.,
  {Wilkinson} M.~I.,  2011, \mn@doi [\aap] {10.1051/0004-6361/201014963}, \href
  {https://ui.adsabs.harvard.edu/abs/2011A&A...525A.153A} {525, A153}

\bibitem[\protect\citeauthoryear{{Agertz} et~al.,}{{Agertz}
  et~al.}{2020}]{2020Agertz}
{Agertz} O.,  et~al., 2020, \mn@doi [\mnras] {10.1093/mnras/stz3053}, \href
  {https://ui.adsabs.harvard.edu/abs/2020MNRAS.491.1656A} {491, 1656}

\bibitem[\protect\citeauthoryear{{Bastian} \& {Lardo}}{{Bastian} \&
  {Lardo}}{2018}]{2018Bastian}
{Bastian} N.,  {Lardo} C.,  2018, \mn@doi [\araa]
  {10.1146/annurev-astro-081817-051839}, \href
  {https://ui.adsabs.harvard.edu/abs/2018ARA&A..56...83B} {56, 83}

\bibitem[\protect\citeauthoryear{{Belokurov} et~al.,}{{Belokurov}
  et~al.}{2006}]{2006_Boo1}
{Belokurov} V.,  et~al., 2006, \mn@doi [\apjl] {10.1086/507324}, \href
  {https://ui.adsabs.harvard.edu/abs/2006ApJ...647L.111B} {647, L111}

\bibitem[\protect\citeauthoryear{{Brauer}, {Ji}, {Frebel}, {Dooley},
  {G{\'o}mez}  \& {O'Shea}}{{Brauer} et~al.}{2019}]{2019Brauer_disrupted_dwarf}
{Brauer} K.,  {Ji} A.~P.,  {Frebel} A.,  {Dooley} G.~A.,  {G{\'o}mez} F.~A.,
  {O'Shea} B.~W.,  2019, \mn@doi [\apj] {10.3847/1538-4357/aafafb}, \href
  {https://ui.adsabs.harvard.edu/abs/2019ApJ...871..247B} {871, 247}

\bibitem[\protect\citeauthoryear{{Brown} et~al.,}{{Brown}
  et~al.}{2014}]{2014Brown_UFDquench}
{Brown} T.~M.,  et~al., 2014, \mn@doi [\apj] {10.1088/0004-637X/796/2/91},
  \href {https://ui.adsabs.harvard.edu/abs/2014ApJ...796...91B} {796, 91}

\bibitem[\protect\citeauthoryear{{Busso}, {Gallino}  \& {Wasserburg}}{{Busso}
  et~al.}{1999}]{Busso1999}
{Busso} M.,  {Gallino} R.,   {Wasserburg} G.~J.,  1999, \mn@doi [\araa]
  {10.1146/annurev.astro.37.1.239}, \href
  {https://ui.adsabs.harvard.edu/abs/1999ARA&A..37..239B} {37, 239}

\bibitem[\protect\citeauthoryear{{Cescutti}, {Fran{\c{c}}ois}, {Matteucci},
  {Cayrel}  \& {Spite}}{{Cescutti} et~al.}{2006}]{2006Cescutti_MW_sprocess}
{Cescutti} G.,  {Fran{\c{c}}ois} P.,  {Matteucci} F.,  {Cayrel} R.,   {Spite}
  M.,  2006, \mn@doi [\aap] {10.1051/0004-6361:20053622}, \href
  {https://ui.adsabs.harvard.edu/abs/2006A&A...448..557C} {448, 557}

\bibitem[\protect\citeauthoryear{{Ceverino}, {Dekel}  \& {Bournaud}}{{Ceverino}
  et~al.}{2010}]{2010Ceverino_Ncells}
{Ceverino} D.,  {Dekel} A.,   {Bournaud} F.,  2010, \mn@doi [\mnras]
  {10.1111/j.1365-2966.2010.16433.x}, \href
  {https://ui.adsabs.harvard.edu/abs/2010MNRAS.404.2151C} {404, 2151}

\bibitem[\protect\citeauthoryear{{Chabrier}}{{Chabrier}}{2001}]{Chabrier01_IMF}
{Chabrier} G.,  2001, \mn@doi [\apj] {10.1086/321401}, \href
  {https://ui.adsabs.harvard.edu/abs/2001ApJ...554.1274C} {554, 1274}

\bibitem[\protect\citeauthoryear{{Chiti}, {Frebel}, {Ji}, {Jerjen}, {Kim}  \&
  {Norris}}{{Chiti} et~al.}{2018}]{2018Chiti_BaFe}
{Chiti} A.,  {Frebel} A.,  {Ji} A.~P.,  {Jerjen} H.,  {Kim} D.,   {Norris}
  J.~E.,  2018, \mn@doi [\apj] {10.3847/1538-4357/aab4fc}, \href
  {https://ui.adsabs.harvard.edu/abs/2018ApJ...857...74C} {857, 74}

\bibitem[\protect\citeauthoryear{{Choplin}, {Hirschi}, {Meynet}  \&
  {Ekstr{\"o}m}}{{Choplin} et~al.}{2017}]{Choplin2017}
{Choplin} A.,  {Hirschi} R.,  {Meynet} G.,   {Ekstr{\"o}m} S.,  2017, \mn@doi
  [\aap] {10.1051/0004-6361/201731948}, \href
  {https://ui.adsabs.harvard.edu/abs/2017A&A...607L...3C} {607, L3}

\bibitem[\protect\citeauthoryear{{Conroy} \& {van Dokkum}}{{Conroy} \& {van
  Dokkum}}{2012}]{2012Conroy}
{Conroy} C.,  {van Dokkum} P.~G.,  2012, \mn@doi [\apj]
  {10.1088/0004-637X/760/1/71}, \href
  {https://ui.adsabs.harvard.edu/abs/2012ApJ...760...71C} {760, 71}

\bibitem[\protect\citeauthoryear{{Cowan}, {Sneden}, {Lawler}, {Aprahamian},
  {Wiescher}, {Langanke}, {Mart{\'\i}nez-Pinedo}  \& {Thielemann}}{{Cowan}
  et~al.}{2021}]{2021Cowan_RprocessReview}
{Cowan} J.~J.,  {Sneden} C.,  {Lawler} J.~E.,  {Aprahamian} A.,  {Wiescher} M.,
   {Langanke} K.,  {Mart{\'\i}nez-Pinedo} G.,   {Thielemann} F.-K.,  2021,
  \mn@doi [Reviews of Modern Physics] {10.1103/RevModPhys.93.015002}, \href
  {https://ui.adsabs.harvard.edu/abs/2021RvMP...93a5002C} {93, 015002}

\bibitem[\protect\citeauthoryear{{Cristallo}, {Straniero}, {Piersanti}  \&
  {Gobrecht}}{{Cristallo} et~al.}{2015}]{2015Cristallo_FRUITY}
{Cristallo} S.,  {Straniero} O.,  {Piersanti} L.,   {Gobrecht} D.,  2015,
  \mn@doi [\apjs] {10.1088/0067-0049/219/2/40}, \href
  {https://ui.adsabs.harvard.edu/abs/2015ApJS..219...40C} {219, 40}

\bibitem[\protect\citeauthoryear{{Cristallo}, {Karinkuzhi}, {Goswami},
  {Piersanti}  \& {Gobrecht}}{{Cristallo} et~al.}{2016}]{2016Cristallo_FRUITY}
{Cristallo} S.,  {Karinkuzhi} D.,  {Goswami} A.,  {Piersanti} L.,   {Gobrecht}
  D.,  2016, \mn@doi [\apj] {10.3847/1538-4357/833/2/181}, \href
  {https://ui.adsabs.harvard.edu/abs/2016ApJ...833..181C} {833, 181}

\bibitem[\protect\citeauthoryear{{Doherty}, {Gil-Pons}, {Siess}  \&
  {Lattanzio}}{{Doherty} et~al.}{2017}]{2017Doherty_superAGB}
{Doherty} C.~L.,  {Gil-Pons} P.,  {Siess} L.,   {Lattanzio} J.~C.,  2017,
  \mn@doi [\pasa] {10.1017/pasa.2017.52}, \href
  {https://ui.adsabs.harvard.edu/abs/2017PASA...34...56D} {34, e056}

\bibitem[\protect\citeauthoryear{{Feltzing}, {Eriksson}, {Kleyna}  \&
  {Wilkinson}}{{Feltzing} et~al.}{2009}]{2009Feltzing_BaFe}
{Feltzing} S.,  {Eriksson} K.,  {Kleyna} J.,   {Wilkinson} M.~I.,  2009,
  \mn@doi [\aap] {10.1051/0004-6361/200912833}, \href
  {https://ui.adsabs.harvard.edu/abs/2009A&A...508L...1F} {508, L1}

\bibitem[\protect\citeauthoryear{{Fran{\c{c}}ois}, {Monaco}, {Bonifacio}, {Moni
  Bidin}, {Geisler}  \& {Sbordone}}{{Fran{\c{c}}ois}
  et~al.}{2016}]{2016Francois_BaFe}
{Fran{\c{c}}ois} P.,  {Monaco} L.,  {Bonifacio} P.,  {Moni Bidin} C.,
  {Geisler} D.,   {Sbordone} L.,  2016, \mn@doi [\aap]
  {10.1051/0004-6361/201527181}, \href
  {https://ui.adsabs.harvard.edu/abs/2016A&A...588A...7F} {588, A7}

\bibitem[\protect\citeauthoryear{{Frebel}, {Simon}, {Geha}  \&
  {Willman}}{{Frebel} et~al.}{2010}]{2010Frebel_BaFe}
{Frebel} A.,  {Simon} J.~D.,  {Geha} M.,   {Willman} B.,  2010, \mn@doi [\apj]
  {10.1088/0004-637X/708/1/560}, \href
  {https://ui.adsabs.harvard.edu/abs/2010ApJ...708..560F} {708, 560}

\bibitem[\protect\citeauthoryear{{Frebel}, {Simon}  \& {Kirby}}{{Frebel}
  et~al.}{2014}]{2014Frebel_BaFe}
{Frebel} A.,  {Simon} J.~D.,   {Kirby} E.~N.,  2014, \mn@doi [\apj]
  {10.1088/0004-637X/786/1/74}, \href
  {https://ui.adsabs.harvard.edu/abs/2014ApJ...786...74F} {786, 74}

\bibitem[\protect\citeauthoryear{{Geha} et~al.,}{{Geha}
  et~al.}{2013}]{2013Geha_shallowIMF}
{Geha} M.,  et~al., 2013, \mn@doi [\apj] {10.1088/0004-637X/771/1/29}, \href
  {https://ui.adsabs.harvard.edu/abs/2013ApJ...771...29G} {771, 29}

\bibitem[\protect\citeauthoryear{{Gennaro} et~al.,}{{Gennaro}
  et~al.}{2018}]{2018_shallow_IMF_in_UFD}
{Gennaro} M.,  et~al., 2018, \mn@doi [\apj] {10.3847/1538-4357/aaa973}, \href
  {https://ui.adsabs.harvard.edu/abs/2018ApJ...855...20G} {855, 20}

\bibitem[\protect\citeauthoryear{{Gilmore}, {Norris}, {Monaco}, {Yong}, {Wyse}
  \& {Geisler}}{{Gilmore} et~al.}{2013}]{2013Gilmore_BaFe}
{Gilmore} G.,  {Norris} J.~E.,  {Monaco} L.,  {Yong} D.,  {Wyse} R. F.~G.,
  {Geisler} D.,  2013, \mn@doi [\apj] {10.1088/0004-637X/763/1/61}, \href
  {https://ui.adsabs.harvard.edu/abs/2013ApJ...763...61G} {763, 61}

\bibitem[\protect\citeauthoryear{{Hahn} \& {Abel}}{{Hahn} \&
  {Abel}}{2011}]{Hahn11_MUSIC}
{Hahn} O.,  {Abel} T.,  2011, \mn@doi [\mnras]
  {10.1111/j.1365-2966.2011.18820.x}, \href
  {https://ui.adsabs.harvard.edu/abs/2011MNRAS.415.2101H} {415, 2101}

\bibitem[\protect\citeauthoryear{{Hansen} et~al.,}{{Hansen}
  et~al.}{2017}]{2017Hansen_TucIII}
{Hansen} T.~T.,  et~al., 2017, \mn@doi [\apj] {10.3847/1538-4357/aa634a}, \href
  {https://ui.adsabs.harvard.edu/abs/2017ApJ...838...44H} {838, 44}

\bibitem[\protect\citeauthoryear{{Hansen} et~al.,}{{Hansen}
  et~al.}{2020}]{2020Hansen_GrusII}
{Hansen} T.~T.,  et~al., 2020, \mn@doi [\apj] {10.3847/1538-4357/ab9643}, \href
  {https://ui.adsabs.harvard.edu/abs/2020ApJ...897..183H} {897, 183}

\bibitem[\protect\citeauthoryear{{Hirai}, {Wanajo}  \& {Saitoh}}{{Hirai}
  et~al.}{2019}]{2019Hirai_Sr_in_dwarfs}
{Hirai} Y.,  {Wanajo} S.,   {Saitoh} T.~R.,  2019, \mn@doi [\apj]
  {10.3847/1538-4357/ab4654}, \href
  {https://ui.adsabs.harvard.edu/abs/2019ApJ...885...33H} {885, 33}

\bibitem[\protect\citeauthoryear{{Hiramatsu} et~al.,}{{Hiramatsu}
  et~al.}{2020}]{2020Hiramatsu_might_be_ECSNe}
{Hiramatsu} D.,  et~al., 2020, arXiv e-prints, \href
  {https://ui.adsabs.harvard.edu/abs/2020arXiv201102176H} {p. arXiv:2011.02176}

\bibitem[\protect\citeauthoryear{{Hirano}, {Hosokawa}, {Yoshida}, {Umeda},
  {Omukai}, {Chiaki}  \& {Yorke}}{{Hirano}
  et~al.}{2014}]{2014Hirano_100firststars}
{Hirano} S.,  {Hosokawa} T.,  {Yoshida} N.,  {Umeda} H.,  {Omukai} K.,
  {Chiaki} G.,   {Yorke} H.~W.,  2014, \mn@doi [\apj]
  {10.1088/0004-637X/781/2/60}, \href
  {https://ui.adsabs.harvard.edu/abs/2014ApJ...781...60H} {781, 60}

\bibitem[\protect\citeauthoryear{{Honda}, {Aoki}, {Kajino}, {Ando}, {Beers},
  {Izumiura}, {Sadakane}  \& {Takada-Hidai}}{{Honda}
  et~al.}{2004}]{2004Honda_hondastar}
{Honda} S.,  {Aoki} W.,  {Kajino} T.,  {Ando} H.,  {Beers} T.~C.,  {Izumiura}
  H.,  {Sadakane} K.,   {Takada-Hidai} M.,  2004, \mn@doi [\apj]
  {10.1086/383406}, \href
  {https://ui.adsabs.harvard.edu/abs/2004ApJ...607..474H} {607, 474}

\bibitem[\protect\citeauthoryear{{Iben}}{{Iben}}{1975}]{Iben1975}
{Iben} I. J.,  1975, \mn@doi [\apj] {10.1086/153433}, \href
  {https://ui.adsabs.harvard.edu/abs/1975ApJ...196..525I} {196, 525}

\bibitem[\protect\citeauthoryear{{Inoue} \& {Yoshida}}{{Inoue} \&
  {Yoshida}}{2019}]{2019Inoue_ISM}
{Inoue} S.,  {Yoshida} N.,  2019, \mn@doi [\mnras] {10.1093/mnras/stz2076},
  \href {https://ui.adsabs.harvard.edu/abs/2019MNRAS.488.4400I} {488, 4400}

\bibitem[\protect\citeauthoryear{{Ishigaki}, {Aoki}, {Arimoto}  \&
  {Okamoto}}{{Ishigaki} et~al.}{2014}]{2014Ishigaki_BaFe}
{Ishigaki} M.~N.,  {Aoki} W.,  {Arimoto} N.,   {Okamoto} S.,  2014, \mn@doi
  [\aap] {10.1051/0004-6361/201322796}, \href
  {https://ui.adsabs.harvard.edu/abs/2014A&A...562A.146I} {562, A146}

\bibitem[\protect\citeauthoryear{{Janka}, {M{\"u}ller}, {Kitaura}  \&
  {Buras}}{{Janka} et~al.}{2008}]{2008Janka_ECSNe}
{Janka} H.~T.,  {M{\"u}ller} B.,  {Kitaura} F.~S.,   {Buras} R.,  2008, \mn@doi
  [\aap] {10.1051/0004-6361:20079334}, \href
  {https://ui.adsabs.harvard.edu/abs/2008A&A...485..199J} {485, 199}

\bibitem[\protect\citeauthoryear{{Jeon}, {Besla}  \& {Bromm}}{{Jeon}
  et~al.}{2017}]{2017Jeon}
{Jeon} M.,  {Besla} G.,   {Bromm} V.,  2017, \mn@doi [\apj]
  {10.3847/1538-4357/aa8c80}, \href
  {https://ui.adsabs.harvard.edu/abs/2017ApJ...848...85J} {848, 85}

\bibitem[\protect\citeauthoryear{{Ji} \& {Frebel}}{{Ji} \&
  {Frebel}}{2018}]{2018Ji_BaFe}
{Ji} A.~P.,  {Frebel} A.,  2018, \mn@doi [\apj] {10.3847/1538-4357/aab14a},
  \href {https://ui.adsabs.harvard.edu/abs/2018ApJ...856..138J} {856, 138}

\bibitem[\protect\citeauthoryear{{Ji}, {Frebel}, {Chiti}  \& {Simon}}{{Ji}
  et~al.}{2016a}]{Ji16_RetII}
{Ji} A.~P.,  {Frebel} A.,  {Chiti} A.,   {Simon} J.~D.,  2016a, \mn@doi [\nat]
  {10.1038/nature17425}, \href
  {https://ui.adsabs.harvard.edu/abs/2016Natur.531..610J} {531, 610}

\bibitem[\protect\citeauthoryear{{Ji}, {Frebel}, {Simon}  \& {Geha}}{{Ji}
  et~al.}{2016b}]{2016Ji_BaFe}
{Ji} A.~P.,  {Frebel} A.,  {Simon} J.~D.,   {Geha} M.,  2016b, \mn@doi [\apj]
  {10.3847/0004-637X/817/1/41}, \href
  {https://ui.adsabs.harvard.edu/abs/2016ApJ...817...41J} {817, 41}

\bibitem[\protect\citeauthoryear{{Ji}, {Frebel}, {Simon}  \& {Chiti}}{{Ji}
  et~al.}{2016c}]{2016Ji_RetII_ApJ}
{Ji} A.~P.,  {Frebel} A.,  {Simon} J.~D.,   {Chiti} A.,  2016c, \mn@doi [\apj]
  {10.3847/0004-637X/830/2/93}, \href
  {https://ui.adsabs.harvard.edu/abs/2016ApJ...830...93J} {830, 93}

\bibitem[\protect\citeauthoryear{{Ji}, {Frebel}, {Ezzeddine}  \& {Casey}}{{Ji}
  et~al.}{2016d}]{2016Ji_BaFe2}
{Ji} A.~P.,  {Frebel} A.,  {Ezzeddine} R.,   {Casey} A.~R.,  2016d, \mn@doi
  [\apjl] {10.3847/2041-8205/832/1/L3}, \href
  {https://ui.adsabs.harvard.edu/abs/2016ApJ...832L...3J} {832, L3}

\bibitem[\protect\citeauthoryear{{Ji}, {Simon}, {Frebel}, {Venn}  \&
  {Hansen}}{{Ji} et~al.}{2019a}]{2019Ji_GCandUFD}
{Ji} A.~P.,  {Simon} J.~D.,  {Frebel} A.,  {Venn} K.~A.,   {Hansen} T.~T.,
  2019a, \mn@doi [\apj] {10.3847/1538-4357/aaf3bb}, \href
  {https://ui.adsabs.harvard.edu/abs/2019ApJ...870...83J} {870, 83}

\bibitem[\protect\citeauthoryear{{Ji}, {Simon}, {Frebel}, {Venn}  \&
  {Hansen}}{{Ji} et~al.}{2019b}]{2019Ji_BaFe}
{Ji} A.~P.,  {Simon} J.~D.,  {Frebel} A.,  {Venn} K.~A.,   {Hansen} T.~T.,
  2019b, \mn@doi [\apj] {10.3847/1538-4357/aaf3bb}, \href
  {https://ui.adsabs.harvard.edu/abs/2019ApJ...870...83J} {870, 83}

\bibitem[\protect\citeauthoryear{{Ji} et~al.,}{{Ji}
  et~al.}{2020}]{2020Ji_CarII}
{Ji} A.~P.,  et~al., 2020, \mn@doi [\apj] {10.3847/1538-4357/ab6213}, \href
  {https://ui.adsabs.harvard.edu/abs/2020ApJ...889...27J} {889, 27}

\bibitem[\protect\citeauthoryear{{K{\"a}ppeler}, {Gallino}, {Bisterzo}  \&
  {Aoki}}{{K{\"a}ppeler} et~al.}{2011}]{2011Kappeler_sprocess_review}
{K{\"a}ppeler} F.,  {Gallino} R.,  {Bisterzo} S.,   {Aoki} W.,  2011, \mn@doi
  [Reviews of Modern Physics] {10.1103/RevModPhys.83.157}, \href
  {https://ui.adsabs.harvard.edu/abs/2011RvMP...83..157K} {83, 157}

\bibitem[\protect\citeauthoryear{{Kirby}, {Cohen}, {Simon}, {Guhathakurta},
  {Thygesen}  \& {Duggan}}{{Kirby} et~al.}{2017}]{2017Kirby_BaFe}
{Kirby} E.~N.,  {Cohen} J.~G.,  {Simon} J.~D.,  {Guhathakurta} P.,  {Thygesen}
  A.~O.,   {Duggan} G.~E.,  2017, \mn@doi [\apj] {10.3847/1538-4357/aa6570},
  \href {https://ui.adsabs.harvard.edu/abs/2017ApJ...838...83K} {838, 83}

\bibitem[\protect\citeauthoryear{{Koch} \& {Rich}}{{Koch} \&
  {Rich}}{2014}]{2014Koch_BaFe}
{Koch} A.,  {Rich} R.~M.,  2014, \mn@doi [\apj] {10.1088/0004-637X/794/1/89},
  \href {https://ui.adsabs.harvard.edu/abs/2014ApJ...794...89K} {794, 89}

\bibitem[\protect\citeauthoryear{{Koch}, {McWilliam}, {Grebel}, {Zucker}  \&
  {Belokurov}}{{Koch} et~al.}{2008}]{2008Koch_BaFe}
{Koch} A.,  {McWilliam} A.,  {Grebel} E.~K.,  {Zucker} D.~B.,   {Belokurov} V.,
   2008, \mn@doi [\apjl] {10.1086/595001}, \href
  {https://ui.adsabs.harvard.edu/abs/2008ApJ...688L..13K} {688, L13}

\bibitem[\protect\citeauthoryear{{Koch} et~al.,}{{Koch}
  et~al.}{2009}]{2009Koch_BaFe}
{Koch} A.,  et~al., 2009, \mn@doi [\apj] {10.1088/0004-637X/690/1/453}, \href
  {https://ui.adsabs.harvard.edu/abs/2009ApJ...690..453K} {690, 453}

\bibitem[\protect\citeauthoryear{{Koch}, {Feltzing}, {Ad{\'e}n}  \&
  {Matteucci}}{{Koch} et~al.}{2013}]{2013Koch_BaFe}
{Koch} A.,  {Feltzing} S.,  {Ad{\'e}n} D.,   {Matteucci} F.,  2013, \mn@doi
  [\aap] {10.1051/0004-6361/201220742}, \href
  {https://ui.adsabs.harvard.edu/abs/2013A&A...554A...5K} {554, A5}

\bibitem[\protect\citeauthoryear{{Komiya}, {Suda}, {Minaguchi}, {Shigeyama},
  {Aoki}  \& {Fujimoto}}{{Komiya} et~al.}{2007}]{2007Komiya}
{Komiya} Y.,  {Suda} T.,  {Minaguchi} H.,  {Shigeyama} T.,  {Aoki} W.,
  {Fujimoto} M.~Y.,  2007, \mn@doi [\apj] {10.1086/510826}, \href
  {https://ui.adsabs.harvard.edu/abs/2007ApJ...658..367K} {658, 367}

\bibitem[\protect\citeauthoryear{{Komiya}, {Suda}  \& {Fujimoto}}{{Komiya}
  et~al.}{2009}]{2009Komiya}
{Komiya} Y.,  {Suda} T.,   {Fujimoto} M.~Y.,  2009, \mn@doi [\apj]
  {10.1088/0004-637X/694/2/1577}, \href
  {https://ui.adsabs.harvard.edu/abs/2009ApJ...694.1577K} {694, 1577}

\bibitem[\protect\citeauthoryear{{Kroupa}, {Weidner}, {Pflamm-Altenburg},
  {Thies}, {Dabringhausen}, {Marks}  \& {Maschberger}}{{Kroupa}
  et~al.}{2013}]{2013Kroupa_IGIMF_book}
{Kroupa} P.,  {Weidner} C.,  {Pflamm-Altenburg} J.,  {Thies} I.,
  {Dabringhausen} J.,  {Marks} M.,   {Maschberger} T.,  2013, {The Stellar and
  Sub-Stellar Initial Mass Function of Simple and Composite Populations}.
p.~115, \mn@doi{10.1007/978-94-007-5612-0_4}

\bibitem[\protect\citeauthoryear{{Lai}, {Lee}, {Bolte}, {Lucatello}, {Beers},
  {Johnson}, {Sivarani}  \& {Rockosi}}{{Lai} et~al.}{2011}]{2011Lai_BaFe}
{Lai} D.~K.,  {Lee} Y.~S.,  {Bolte} M.,  {Lucatello} S.,  {Beers} T.~C.,
  {Johnson} J.~A.,  {Sivarani} T.,   {Rockosi} C.~M.,  2011, \mn@doi [\apj]
  {10.1088/0004-637X/738/1/51}, \href
  {https://ui.adsabs.harvard.edu/abs/2011ApJ...738...51L} {738, 51}

\bibitem[\protect\citeauthoryear{{Lee}, {Suda}, {Beers}  \& {Stancliffe}}{{Lee}
  et~al.}{2014}]{2014Lee_IMF}
{Lee} Y.~S.,  {Suda} T.,  {Beers} T.~C.,   {Stancliffe} R.~J.,  2014, \mn@doi
  [\apj] {10.1088/0004-637X/788/2/131}, \href
  {https://ui.adsabs.harvard.edu/abs/2014ApJ...788..131L} {788, 131}

\bibitem[\protect\citeauthoryear{{Limongi} \& {Chieffi}}{{Limongi} \&
  {Chieffi}}{2018}]{2018Limongi_Chieffi_RMS}
{Limongi} M.,  {Chieffi} A.,  2018, \mn@doi [\apjs] {10.3847/1538-4365/aacb24},
  \href {https://ui.adsabs.harvard.edu/abs/2018ApJS..237...13L} {237, 13}

\bibitem[\protect\citeauthoryear{{Maoz}, {Mannucci}  \& {Brandt}}{{Maoz}
  et~al.}{2012}]{2012Maoz_Ia}
{Maoz} D.,  {Mannucci} F.,   {Brandt} T.~D.,  2012, \mn@doi [\mnras]
  {10.1111/j.1365-2966.2012.21871.x}, \href
  {https://ui.adsabs.harvard.edu/abs/2012MNRAS.426.3282M} {426, 3282}

\bibitem[\protect\citeauthoryear{{Marshall} et~al.,}{{Marshall}
  et~al.}{2019}]{Marshall18_TucIIIobservation}
{Marshall} J.~L.,  et~al., 2019, \mn@doi [\apj] {10.3847/1538-4357/ab3653},
  \href {https://ui.adsabs.harvard.edu/abs/2019ApJ...882..177M} {882, 177}

\bibitem[\protect\citeauthoryear{{Martin}, {Ibata}, {Chapman}, {Irwin}  \&
  {Lewis}}{{Martin} et~al.}{2007}]{2007Martin_BaFe}
{Martin} N.~F.,  {Ibata} R.~A.,  {Chapman} S.~C.,  {Irwin} M.,   {Lewis} G.~F.,
   2007, \mn@doi [\mnras] {10.1111/j.1365-2966.2007.12055.x}, \href
  {https://ui.adsabs.harvard.edu/abs/2007MNRAS.380..281M} {380, 281}

\bibitem[\protect\citeauthoryear{{Martin} et~al.,}{{Martin}
  et~al.}{2016}]{2016Martin_BaFe}
{Martin} N.~F.,  et~al., 2016, \mn@doi [\apj] {10.3847/0004-637X/818/1/40},
  \href {https://ui.adsabs.harvard.edu/abs/2016ApJ...818...40M} {818, 40}

\bibitem[\protect\citeauthoryear{{Matsuno} et~al.,}{{Matsuno}
  et~al.}{2020}]{2020Matsuno_no_sprocess}
{Matsuno} T.,  et~al., 2020, arXiv e-prints, \href
  {https://ui.adsabs.harvard.edu/abs/2020arXiv200603619M} {p. arXiv:2006.03619}

\bibitem[\protect\citeauthoryear{{Meynet}, {Ekstr{\"o}m}  \& {Maeder}}{{Meynet}
  et~al.}{2006}]{2006Meynet_rotation_CNOcycle}
{Meynet} G.,  {Ekstr{\"o}m} S.,   {Maeder} A.,  2006, \mn@doi [\aap]
  {10.1051/0004-6361:20053070}, \href
  {https://ui.adsabs.harvard.edu/abs/2006A&A...447..623M} {447, 623}

\bibitem[\protect\citeauthoryear{{Montes} et~al.,}{{Montes}
  et~al.}{2007}]{2007Montes_LEPP}
{Montes} F.,  et~al., 2007, \mn@doi [\apj] {10.1086/523084}, \href
  {https://ui.adsabs.harvard.edu/abs/2007ApJ...671.1685M} {671, 1685}

\bibitem[\protect\citeauthoryear{{Nagasawa} et~al.,}{{Nagasawa}
  et~al.}{2018}]{2018Nagasawa_BaFe}
{Nagasawa} D.~Q.,  et~al., 2018, \mn@doi [\apj] {10.3847/1538-4357/aaa01d},
  \href {https://ui.adsabs.harvard.edu/abs/2018ApJ...852...99N} {852, 99}

\bibitem[\protect\citeauthoryear{{Nishimura}, {Sawai}, {Takiwaki}, {Yamada}  \&
  {Thielemann}}{{Nishimura} et~al.}{2017}]{Nishimura17_iprocess}
{Nishimura} N.,  {Sawai} H.,  {Takiwaki} T.,  {Yamada} S.,   {Thielemann}
  F.~K.,  2017, \mn@doi [\apjl] {10.3847/2041-8213/aa5dee}, \href
  {https://ui.adsabs.harvard.edu/abs/2017ApJ...836L..21N} {836, L21}

\bibitem[\protect\citeauthoryear{{Nomoto}}{{Nomoto}}{1987}]{1987Nomoto_ECSNe}
{Nomoto} K.,  1987, \mn@doi [\apj] {10.1086/165716}, \href
  {https://ui.adsabs.harvard.edu/abs/1987ApJ...322..206N} {322, 206}

\bibitem[\protect\citeauthoryear{{Norris}, {Yong}, {Gilmore}  \&
  {Wyse}}{{Norris} et~al.}{2010a}]{2010Norris_BaFe2}
{Norris} J.~E.,  {Yong} D.,  {Gilmore} G.,   {Wyse} R. F.~G.,  2010a, \mn@doi
  [\apj] {10.1088/0004-637X/711/1/350}, \href
  {https://ui.adsabs.harvard.edu/abs/2010ApJ...711..350N} {711, 350}

\bibitem[\protect\citeauthoryear{{Norris}, {Gilmore}, {Wyse}, {Yong}  \&
  {Frebel}}{{Norris} et~al.}{2010b}]{2010Norris_BaFe3}
{Norris} J.~E.,  {Gilmore} G.,  {Wyse} R. F.~G.,  {Yong} D.,   {Frebel} A.,
  2010b, \mn@doi [\apjl] {10.1088/2041-8205/722/1/L104}, \href
  {https://ui.adsabs.harvard.edu/abs/2010ApJ...722L.104N} {722, L104}

\bibitem[\protect\citeauthoryear{{Norris}, {Wyse}, {Gilmore}, {Yong}, {Frebel},
  {Wilkinson}, {Belokurov}  \& {Zucker}}{{Norris}
  et~al.}{2010c}]{2010Norris_BaFe1}
{Norris} J.~E.,  {Wyse} R. F.~G.,  {Gilmore} G.,  {Yong} D.,  {Frebel} A.,
  {Wilkinson} M.~I.,  {Belokurov} V.,   {Zucker} D.~B.,  2010c, \mn@doi [\apj]
  {10.1088/0004-637X/723/2/1632}, \href
  {https://ui.adsabs.harvard.edu/abs/2010ApJ...723.1632N} {723, 1632}

\bibitem[\protect\citeauthoryear{{Pakmor}, {Springel}, {Bauer}, {Mocz},
  {Munoz}, {Ohlmann}, {Schaal}  \& {Zhu}}{{Pakmor}
  et~al.}{2016}]{Pakmor16_AREPO_improvement}
{Pakmor} R.,  {Springel} V.,  {Bauer} A.,  {Mocz} P.,  {Munoz} D.~J.,
  {Ohlmann} S.~T.,  {Schaal} K.,   {Zhu} C.,  2016, \mn@doi [\mnras]
  {10.1093/mnras/stv2380}, \href
  {https://ui.adsabs.harvard.edu/abs/2016MNRAS.455.1134P} {455, 1134}

\bibitem[\protect\citeauthoryear{{Planck Collaboration} et~al.,}{{Planck
  Collaboration} et~al.}{2020}]{Planck2018}
{Planck Collaboration} et~al., 2020, \mn@doi [\aap]
  {10.1051/0004-6361/201833910}, \href
  {https://ui.adsabs.harvard.edu/abs/2020A&A...641A...6P} {641, A6}

\bibitem[\protect\citeauthoryear{{Portinari}, {Chiosi}  \&
  {Bressan}}{{Portinari} et~al.}{1998}]{1998Portinari_typeIIyield}
{Portinari} L.,  {Chiosi} C.,   {Bressan} A.,  1998, \aap, \href
  {https://ui.adsabs.harvard.edu/abs/1998A&A...334..505P} {334, 505}

\bibitem[\protect\citeauthoryear{{Prantzos}, {Abia}, {Limongi}, {Chieffi}  \&
  {Cristallo}}{{Prantzos} et~al.}{2018}]{2018Prantzos_RMS}
{Prantzos} N.,  {Abia} C.,  {Limongi} M.,  {Chieffi} A.,   {Cristallo} S.,
  2018, \mn@doi [\mnras] {10.1093/mnras/sty316}, \href
  {https://ui.adsabs.harvard.edu/abs/2018MNRAS.476.3432P} {476, 3432}

\bibitem[\protect\citeauthoryear{{Renzini}}{{Renzini}}{2008}]{2008Renzini_MassBudgetProblem}
{Renzini} A.,  2008, \mn@doi [\mnras] {10.1111/j.1365-2966.2008.13892.x}, \href
  {https://ui.adsabs.harvard.edu/abs/2008MNRAS.391..354R} {391, 354}

\bibitem[\protect\citeauthoryear{{Roederer} et~al.,}{{Roederer}
  et~al.}{2016}]{2016Roederer_BaFe}
{Roederer} I.~U.,  et~al., 2016, \mn@doi [\aj] {10.3847/0004-6256/151/3/82},
  \href {https://ui.adsabs.harvard.edu/abs/2016AJ....151...82R} {151, 82}

\bibitem[\protect\citeauthoryear{{Safarzadeh}, {Ji}, {Dooley}, {Frebel},
  {Scannapieco}, {G{\'o}mez}  \& {O'Shea}}{{Safarzadeh}
  et~al.}{2018}]{Safarzadeh18_UFDselection}
{Safarzadeh} M.,  {Ji} A.~P.,  {Dooley} G.~A.,  {Frebel} A.,  {Scannapieco} E.,
   {G{\'o}mez} F.~A.,   {O'Shea} B.~W.,  2018, \mn@doi [\mnras]
  {10.1093/mnras/sty595}, \href
  {https://ui.adsabs.harvard.edu/abs/2018MNRAS.476.5006S} {476, 5006}

\bibitem[\protect\citeauthoryear{{Salpeter}}{{Salpeter}}{1955}]{1955_Salpeter}
{Salpeter} E.~E.,  1955, \mn@doi [\apj] {10.1086/145971}, \href
  {https://ui.adsabs.harvard.edu/abs/1955ApJ...121..161S} {121, 161}

\bibitem[\protect\citeauthoryear{{Simon}}{{Simon}}{2018}]{2018Simon_tidalstripping}
{Simon} J.~D.,  2018, \mn@doi [\apj] {10.3847/1538-4357/aacdfb}, \href
  {https://ui.adsabs.harvard.edu/abs/2018ApJ...863...89S} {863, 89}

\bibitem[\protect\citeauthoryear{{Simon}}{{Simon}}{2019}]{Simon19_UFDreview}
{Simon} J.~D.,  2019, \mn@doi [\araa] {10.1146/annurev-astro-091918-104453},
  \href {https://ui.adsabs.harvard.edu/abs/2019ARA&A..57..375S} {57, 375}

\bibitem[\protect\citeauthoryear{{Simon} et~al.,}{{Simon}
  et~al.}{2011}]{2011Simon_BaFe}
{Simon} J.~D.,  et~al., 2011, \mn@doi [\apj] {10.1088/0004-637X/733/1/46},
  \href {https://ui.adsabs.harvard.edu/abs/2011ApJ...733...46S} {733, 46}

\bibitem[\protect\citeauthoryear{{Sneden}, {Cowan}  \& {Gallino}}{{Sneden}
  et~al.}{2008}]{2008Sneden_review}
{Sneden} C.,  {Cowan} J.~J.,   {Gallino} R.,  2008, \mn@doi [\araa]
  {10.1146/annurev.astro.46.060407.145207}, \href
  {https://ui.adsabs.harvard.edu/abs/2008ARA&A..46..241S} {46, 241}

\bibitem[\protect\citeauthoryear{{Sneden}, {Lawler}, {Cowan}, {Ivans}  \& {Den
  Hartog}}{{Sneden} et~al.}{2009}]{2009Sneden}
{Sneden} C.,  {Lawler} J.~E.,  {Cowan} J.~J.,  {Ivans} I.~I.,   {Den Hartog}
  E.~A.,  2009, \mn@doi [\apjs] {10.1088/0067-0049/182/1/80}, \href
  {https://ui.adsabs.harvard.edu/abs/2009ApJS..182...80S} {182, 80}

\bibitem[\protect\citeauthoryear{{Springel}}{{Springel}}{2010}]{Springel10_AREPO}
{Springel} V.,  2010, \mn@doi [\mnras] {10.1111/j.1365-2966.2009.15715.x},
  \href {https://ui.adsabs.harvard.edu/abs/2010MNRAS.401..791S} {401, 791}

\bibitem[\protect\citeauthoryear{{Springel} \& {Hernquist}}{{Springel} \&
  {Hernquist}}{2003}]{Springel03_WindParticles}
{Springel} V.,  {Hernquist} L.,  2003, \mn@doi [\mnras]
  {10.1046/j.1365-8711.2003.06206.x}, \href
  {https://ui.adsabs.harvard.edu/abs/2003MNRAS.339..289S} {339, 289}

\bibitem[\protect\citeauthoryear{{Suda}, {Aikawa}, {Machida}, {Fujimoto}  \&
  {Iben}}{{Suda} et~al.}{2004}]{Suda2004}
{Suda} T.,  {Aikawa} M.,  {Machida} M.~N.,  {Fujimoto} M.~Y.,   {Iben} Icko J.,
   2004, \mn@doi [\apj] {10.1086/422135}, \href
  {https://ui.adsabs.harvard.edu/abs/2004ApJ...611..476S} {611, 476}

\bibitem[\protect\citeauthoryear{{Suda} et~al.,}{{Suda}
  et~al.}{2008}]{2008SAGA}
{Suda} T.,  et~al., 2008, \mn@doi [\pasj] {10.1093/pasj/60.5.1159}, \href
  {https://ui.adsabs.harvard.edu/abs/2008PASJ...60.1159S} {60, 1159}

\bibitem[\protect\citeauthoryear{{Suda} et~al.,}{{Suda}
  et~al.}{2013}]{2013Suda_IMF}
{Suda} T.,  et~al., 2013, \mn@doi [\mnras] {10.1093/mnrasl/slt033}, \href
  {https://ui.adsabs.harvard.edu/abs/2013MNRAS.432L..46S} {432, L46}

\bibitem[\protect\citeauthoryear{{Suda} et~al.,}{{Suda}
  et~al.}{2017}]{SAGA_dwarf}
{Suda} T.,  et~al., 2017, \mn@doi [\pasj] {10.1093/pasj/psx059}, \href
  {https://ui.adsabs.harvard.edu/abs/2017PASJ...69...76S} {69, 76}

\bibitem[\protect\citeauthoryear{{Tarumi}, {Yoshida}  \& {Inoue}}{{Tarumi}
  et~al.}{2020}]{2020Tarumi}
{Tarumi} Y.,  {Yoshida} N.,   {Inoue} S.,  2020, \mn@doi [\mnras]
  {10.1093/mnras/staa720}, \href
  {https://ui.adsabs.harvard.edu/abs/2020MNRAS.494..120T} {494, 120}

\bibitem[\protect\citeauthoryear{{Tarumi}, {Yoshida}  \& {Frebel}}{{Tarumi}
  et~al.}{2021}]{2021Tarumi_UFDmerger}
{Tarumi} Y.,  {Yoshida} N.,   {Frebel} A.,  2021, arXiv e-prints, \href
  {https://ui.adsabs.harvard.edu/abs/2021arXiv210301962T} {p. arXiv:2103.01962}

\bibitem[\protect\citeauthoryear{{Travaglio}, {Hillebrandt}, {Reinecke}  \&
  {Thielemann}}{{Travaglio} et~al.}{2004a}]{2004Travaglio_typeIayield}
{Travaglio} C.,  {Hillebrandt} W.,  {Reinecke} M.,   {Thielemann} F.~K.,
  2004a, \mn@doi [\aap] {10.1051/0004-6361:20041108}, \href
  {https://ui.adsabs.harvard.edu/abs/2004A&A...425.1029T} {425, 1029}

\bibitem[\protect\citeauthoryear{{Travaglio}, {Gallino}, {Arnone}, {Cowan},
  {Jordan}  \& {Sneden}}{{Travaglio} et~al.}{2004b}]{2004Travaglio_sprocess}
{Travaglio} C.,  {Gallino} R.,  {Arnone} E.,  {Cowan} J.,  {Jordan} F.,
  {Sneden} C.,  2004b, \mn@doi [\apj] {10.1086/380507}, \href
  {https://ui.adsabs.harvard.edu/abs/2004ApJ...601..864T} {601, 864}

\bibitem[\protect\citeauthoryear{{Truelove}, {Klein}, {McKee}, {Holliman},
  {Howell}  \& {Greenough}}{{Truelove}
  et~al.}{1997}]{1997Truelove_TrueloveCondition}
{Truelove} J.~K.,  {Klein} R.~I.,  {McKee} C.~F.,  {Holliman} John~H. I.,
  {Howell} L.~H.,   {Greenough} J.~A.,  1997, \mn@doi [\apjl] {10.1086/310975},
  \href {https://ui.adsabs.harvard.edu/abs/1997ApJ...489L.179T} {489, L179}

\bibitem[\protect\citeauthoryear{{Vargas}, {Geha}, {Kirby}  \&
  {Simon}}{{Vargas} et~al.}{2013}]{2013Vargas_BaFe}
{Vargas} L.~C.,  {Geha} M.,  {Kirby} E.~N.,   {Simon} J.~D.,  2013, \mn@doi
  [\apj] {10.1088/0004-637X/767/2/134}, \href
  {https://ui.adsabs.harvard.edu/abs/2013ApJ...767..134V} {767, 134}

\bibitem[\protect\citeauthoryear{{Venn}, {Starkenburg}, {Malo}, {Martin}  \&
  {Laevens}}{{Venn} et~al.}{2017}]{2017Venn_BaFe}
{Venn} K.~A.,  {Starkenburg} E.,  {Malo} L.,  {Martin} N.,   {Laevens}
  B.~P.~M.,  2017, \mn@doi [\mnras] {10.1093/mnras/stw3198}, \href
  {https://ui.adsabs.harvard.edu/abs/2017MNRAS.466.3741V} {466, 3741}

\bibitem[\protect\citeauthoryear{{Vogelsberger}, {Genel}, {Sijacki}, {Torrey},
  {Springel}  \& {Hernquist}}{{Vogelsberger}
  et~al.}{2013}]{Vogelsberger13_MetalStripping}
{Vogelsberger} M.,  {Genel} S.,  {Sijacki} D.,  {Torrey} P.,  {Springel} V.,
  {Hernquist} L.,  2013, \mn@doi [\mnras] {10.1093/mnras/stt1789}, \href
  {https://ui.adsabs.harvard.edu/abs/2013MNRAS.436.3031V} {436, 3031}

\bibitem[\protect\citeauthoryear{{Wanajo}, {Janka}  \& {M{\"u}ller}}{{Wanajo}
  et~al.}{2011}]{2011Wanajo_ECSNe}
{Wanajo} S.,  {Janka} H.-T.,   {M{\"u}ller} B.,  2011, \mn@doi [\apjl]
  {10.1088/2041-8205/726/2/L15}, \href
  {https://ui.adsabs.harvard.edu/abs/2011ApJ...726L..15W} {726, L15}

\bibitem[\protect\citeauthoryear{{Wanajo}, {M{\"u}ller}, {Janka}  \&
  {Heger}}{{Wanajo} et~al.}{2018}]{2018Wanajo_ECSNe}
{Wanajo} S.,  {M{\"u}ller} B.,  {Janka} H.-T.,   {Heger} A.,  2018, \mn@doi
  [\apj] {10.3847/1538-4357/aa9d97}, \href
  {https://ui.adsabs.harvard.edu/abs/2018ApJ...852...40W} {852, 40}

\bibitem[\protect\citeauthoryear{{Weinberger}, {Springel}  \&
  {Pakmor}}{{Weinberger} et~al.}{2020}]{2019AREPOrelease}
{Weinberger} R.,  {Springel} V.,   {Pakmor} R.,  2020, \mn@doi [\apjs]
  {10.3847/1538-4365/ab908c}, \href
  {https://ui.adsabs.harvard.edu/abs/2020ApJS..248...32W} {248, 32}

\bibitem[\protect\citeauthoryear{{Wheeler} et~al.,}{{Wheeler}
  et~al.}{2019}]{2019Wheeler}
{Wheeler} C.,  et~al., 2019, \mn@doi [\mnras] {10.1093/mnras/stz2887}, \href
  {https://ui.adsabs.harvard.edu/abs/2019MNRAS.490.4447W} {490, 4447}

\bibitem[\protect\citeauthoryear{{Zucker} et~al.,}{{Zucker}
  et~al.}{2006}]{2006_UMa2}
{Zucker} D.~B.,  et~al., 2006, \mn@doi [\apjl] {10.1086/508628}, \href
  {https://ui.adsabs.harvard.edu/abs/2006ApJ...650L..41Z} {650, L41}

\makeatother
\end{thebibliography}

%%%%%%%%%%%%%%%%%%%%%%%%%%%%%%%%%%%%%%%%%%%%%%%%%%

%%%%%%%%%%%%%%%%%%%%%%%%%%%%%%%%%%%%%%%%%%%%%%%%%%

% Don't change these lines
\bsp	% typesetting comment
\label{lastpage}
\end{document}